\documentclass[a4paper,11pt]{article}
\usepackage{aaskaiid}

\usepackage{subcaption}
\usepackage{graphicx}
\usepackage{orcidlink}

\title{Radio Halos in Galaxy Clusters as unveiled by the SKA telescope}
\ShortTitle{Radio Halos}
\author[1]{R. Cassano\orcidlink{0000-0003-4046-0637}}
\author[1]{G. Di Gennaro\orcidlink{0000-0002-8648-8507}}
\author[3,1]{V. Cuciti\orcidlink{0000-0003-4454-132X}}
\author[2]{A. Datta\orcidlink{0000-0002-5333-1095}}
\author[3,10]{M. Balboni\orcidlink{0009-0001-3048-0020}}
\author[1]{G. Bernardi\orcidlink{0000-0002-0916-7443}}
\author[3,1]{A. Bonafede\orcidlink{0000-0002-5068-4581}}
\author[1]{A. Botteon\orcidlink{0000-0002-9325-1567}}
\author[4]{M. Br{\"u}ggen\orcidlink{0000-0002-3369-7735}}
\author[1]{G. Brunetti\orcidlink{0000-0003-4195-8613}}
\author[5]{S. Chatterjee\orcidlink{0000-0001-8194-8714}}
\author[6,7]{K. Dolag\orcidlink{0000-0003-1750-286X}}
\author[8,9]{S. Ettori\orcidlink{0000-0003-4117-8617}}
\author[10]{F. Gastaldello\orcidlink{0000-0002-9112-0184}}
\author[11]{S. Giacintucci\orcidlink{0000-0002-1634-9886}}
\author[8,9]{C. Giocoli\orcidlink{0000-0002-9590-7961}}
\author[3,1]{M. Gitti\orcidlink{0000-0002-0843-3009}}
\author[12]{R. Kale\orcidlink{0000-0003-1449-3718}}
\author[13]{M. Pandey-Pommier\orcidlink{0000-0001-5829-1099}}
\author[14]{G.~W. Pratt}
\author[15]{M. Rahaman\orcidlink{0000-0002-1372-6017}}
\author[10]{M. Rossetti\orcidlink{0000-0002-9775-732X}}
\author[16]{H. J. A. R\"ottgering\orcidlink{0000-0001-8887-2257}}
\author[12]{R. Santra\orcidlink{0009-0002-0373-570X}}
\author[1,3]{K. S. L. Srikanth\orcidlink{0009-0009-1658-1405}}
\author[16]{R.~J. van Weeren\orcidlink{0000-0002-0587-1660}}
\author[1]{T. Venturi\orcidlink{0000-0002-8476-6307}}

\ShortName{Cassano et al.} 

\affiliation[1]{Istituto Nazionale di Astrofisica (INAF) -- Istituto di
Radioastronomia (IRA), via Gobetti 101, 40129 Bologna, Italy}
\emailAdd{rossella.cassano@inaf.it}
\affiliation[2]{Department of Astronomy Astrophysics and Space Engineering, Indian Institute of Technology Indore, Khandwa Road, Simrol, Indore, Madhya Pradesh, 453552, India}
\affiliation[3]{Dipartimento di Fisica e Astronomia (DIFA), Universit{\`a} di Bologna, via Gobetti 93/2, 40129 Bologna, Italy}
\affiliation[4]{Hamburger Sternwarte, Universit{\"a}t Hamburg, Gojenbergsweg 112, D-21029 Hamburg, Germany}
\affiliation[5]{Centre for Radio Astronomy Techniques and Technologies, Department of Physics and Electronics, Rhodes University, Artillery Road, Makhanda 6139, South Africa}
\affiliation[8]{Istituto Nazionale di Astrofisica (INAF) -- Osservatorio di Astrofisica e Scienza dello Spazio, via Piero Gobetti 93/3, 40129 Bologna, Italy}
\affiliation[9]{INFN, Sezione di Bologna, viale Berti Pichat 6/2, 40127 Bologna, Italy}
\affiliation[6]{Universit{\"a}t-Sternwarte, Fakult{\"a}t f{\"u}r Physik, Ludwig-Maximilians-Universit{\"a}t M{\"u}nchen, Scheinerstr.1, 81679}
\affiliation[7]{Max-Plank-Institut für Astrophysik, Karl-Schwarzschild Strasse 1, D-85740 Garching, Germany}
\affiliation[10]{Istituto Nazionale di Astrofisica – Istituto di Astrofisica Spaziale e Fisica cosmica (IASF), Via A. Corti 12, 20133 Milano, Italy}
\affiliation[11]{Naval Research Laboratory, 
4555 Overlook Avenue SW, Code 7213, 
Washington, DC 20375, USA}
\affiliation[12]{National Centre for Radio Astrophysics, Tata Institute of Fundamental Research, S. P. Pune University Campus, Ganeshkhind, Pune 411007, India}
\affiliation[13]{Pole Scientific, University Catholic of Lyon- University of Lyon, 10 place des Archives 69288, Lyon, France}
\affiliation[14]{Universit\'e Paris-Saclay, Universit\'e Paris Cit\'e, CEA, CNRS, AIM de Paris-Saclay, 91191 Gif-sur-Yvette, France}
\affiliation[15]{Institute of Astronomy, National Tsing Hua University, Hsinchu 300044, Taiwan}
\affiliation[16]{Leiden Observatory, Leiden University, PO Box 9513, NL-2300 RA Leiden, The Netherlands}

\def \eg {e.g.}

\def \ie {i.e.}

\abstract{
Giant radio halos (RHs) are diffuse, Mpc-scale synchrotron sources observed in a growing fraction of galaxy clusters. They trace relativistic particles and magnetic fields in the intracluster medium (ICM), providing a unique window into non-thermal processes and their role in cluster evolution. RHs are primarily found in merging systems, supporting models in which turbulence generated during cluster collisions re-accelerates pre-existing electrons to the energies required for the observed radio emission. In this scenario, the occurrence, power, and spectral properties of RHs depend on the energetics of cluster mergers, with the most massive and dynamically disturbed clusters hosting the most powerful halos. Low-frequency observations are crucial to uncover ultra-steep-spectrum RHs, a key prediction of turbulent re-acceleration models, and are expected to arise from less energetic merger events. LOFAR has enabled statistical studies of large cluster samples, placing robust constraints on RH occurrence and spectral trends. In this Chapter, we model RH formation and evolution using Monte Carlo simulations calibrated on LoTSS-DR2 findings, and we present predictions for SKA-Low in the AA4 configuration. Our results show that SKA will probe an unprecedented region of cluster mass and redshift space, detecting at least $\sim 2500$ RHs up to $z \approx 0.6$, including $\gtrsim 1000$ ultra-steep-spectrum systems, and revealing halos in clusters down to $\sim 10^{14}\, M_\odot$ and out to $z \approx 1$. These surveys will provide stringent tests of turbulent re-acceleration models and significantly advance our understanding of non-thermal processes in galaxy clusters.}

\begin{document}
\maketitle

\section{Introduction}
\label{Sect.intro}

Giant Radio Halos (RHs) are diffuse, Mpc-scale synchrotron radio sources with steep spectra ($\alpha > 1$, with $f_\nu \propto \nu^{-\alpha}$) observed in the central regions of a fraction of galaxy clusters \citep[e.g.][]{feretti2012,vanweeren2019,cassano2023}. They trace the energy content and distribution of relativistic particles and magnetic fields in the intracluster medium (ICM), thus providing unique insights into the non-thermal components of clusters and their role in cluster formation and evolution.

Although RHs are predominantly found in dynamically disturbed, merging clusters—consistent with the idea that turbulence generated during mergers powers particle re-acceleration \citep{brunetti2014,vanweeren2019}—recent observational results are revealing a more complex picture. A handful of apparently relaxed systems host large-scale diffuse emission \citep[e. g.][]{2019MNRAS.486L..80K}. For instance, the cool-core cluster CL1821+643 contains a giant RH despite its regular X-ray morphology \citep{bonafede2014,savini2018}. Similarly, deep LOFAR observations of the Perseus cluster uncovered Mpc-scale diffuse emission well extending beyond the central mini-halo \citep{vanweeren24pers}. However, weak-lensing analyses have revealed evidence of a past merger in Perseus \citep{kim25}, suggesting that low-contrast or older merger events—undetectable through traditional X-ray morphology alone—may still inject turbulence sufficient to sustain cluster-wide synchrotron emission. Also, a statistical study with LOFAR (see also Sect.\ref{Sect.LOFAR_advances}) shows that a non-negligible fraction of clusters classified as more relaxed exhibit RH-like emission \citep{cassano+23}.  

According to the turbulent re-acceleration scenario, turbulence generated during cluster mergers re-energizes pre-existing populations of relativistic electrons—either fossil or secondary—to the energies required to produce the observed synchrotron emission \citep{brunetti2001,petrosian2001,brunetti2016}. In these models, the occurrence and spectral properties of RHs depend on both the merger history and the mass of the host cluster, which determine the available turbulent energy budget. A key prediction is that RHs preferentially form in massive, energetic mergers, whereas they are rarer in less massive systems, where the weaker turbulence leads to very steep radio spectra that become under-luminous at GHz frequencies \citep{cassano2006,cassano10lofar,donnert2013,brunettivazza20}. 
A possible way to incorporate within this framework the emerging evidence of RHs in less dynamically disturbed systems is to assume that these halos are produced during less significant dynamical events, which generate weaker perturbations in the ICM and, consequently, lower levels of turbulent energy. 
Future XRISM~Resolve observations will be essential to test whether weak or steep-spectrum radio halos are powered by modest ICM turbulence. Current measurements indicate generally low small-scale turbulence \citep[$\sigma_v \approx 115$--$170~\mathrm{km\,s^{-1}}$; non-thermal pressure fractions of a few percent;][]{XRISM-Perseus_2025, XRISM_A2029_2025, XRISM_2319_2025}. Cosmological simulations, however, suggest that these can be explained by spatially localized and patchy turbulent regions, without requiring extremely steep turbulence spectra \citep{Vazza_2025}.

This theoretical framework is also consistent with the observed radio bimodality between radio-loud and radio-quiet clusters \citep{cassano2013,2015A&A...579A..92K,cuciti21b,cuciti23} and is supported by the discovery of ultra-steep-spectrum halos \citep{brunetti08nature,dallacasa09,digennaro+21b,santra2024deep}.

An additional contribution to diffuse cluster-scale emission may arise from secondary electrons produced in hadronic collisions between cosmic-ray (CR) protons and thermal ions in the ICM \citep{dennison1980,blasi1999}. Although $\gamma$-ray observations constrain this mechanism to be sub-dominant in the generation of classical RHs \citep{ackermann2016,brunetti2017,adam2021}, it may still contribute to the formation of fainter, ``off-state'' halos in more relaxed systems \citep{brunetti2011,brown2011,donnert2013,cassano2015}.

Recent LOFAR surveys, in particular the LoTSS-DR2 \citep{Shimwell2022}, have dramatically expanded the sample of known RHs \citep{botteon2022}, allowing the study of their radio flux density and redshift distributions, which have been found to be in line with the expectations of the re-acceleration scenario \citep{cassano2023}. It was also found that the fraction of clusters with radio halos increases with the cluster mass, confirming the leading role of the gravitational process of cluster formation in the generation of radio halos. Furthermore, the combination of CHEX–MATE observations \citep{CHEX-MATE21}, based on a homogeneous and deep XMM-Newton follow-up of hundreds of SZ-selected clusters, with LoTSS-DR2 and MeerKAT data has opened the way to spatially resolved studies of the connection between the thermal and non-thermal components of the ICM. These works \citep{Balboni24,Balboni25l} enable both global and spatially resolved analyses of radio-halo properties and their dependence on cluster mass, redshift, and dynamical state, showing that more disturbed systems tend to host brighter and more extended halos than relaxed clusters. In addition, MeerKAT follow-ups of CHEX–MATE clusters allow the detection of new radio halos \citep{Balboni25m}, supporting the radio power–mass relation and suggesting that radio-halo emissivity is the main driver of this correlation.


In this Chapter, we adopt the turbulent re-acceleration scenario as constrained by the LoTSS-DR2 study of Planck clusters to model the expected population of RHs across cosmic time. Using Monte Carlo simulations that link cluster merger histories, magnetic fields, and particle acceleration, we predict the occurrence, luminosity functions, and spectral properties of RHs in the context of forthcoming SKA-Low surveys. In particular, we explore the discovery potential of SKA-Low in AA4 configuration,
whose expected sensitivities ($\sim$20 $\mu$Jy/beam at 150 MHz at a resolution of 10 arcsec \citealt{braun2019}) will enable the detection and characterization of the full population of RHs, including faint and ultra-steep-spectrum systems. These observations, combined with ongoing LOFAR and MeerKAT, and future SKA-Mid observations, will open a new window on the non-thermal life cycle of galaxy clusters and on the physics of turbulent particle acceleration in the ICM.

\noindent A $\Lambda$CDM cosmology ($H_{0}=70\,\rm km\,\rm s^{-1}\,\rm Mpc^{-1}$, $\Omega_{m}=0.3$, $\Omega_{\Lambda}=0.7$) is adopted.

\begin{figure}[t]
\centering
\begin{subfigure}[t]{0.4\textwidth} 
    \centering
    \includegraphics[width=\linewidth,trim=0 0 0 0,clip]{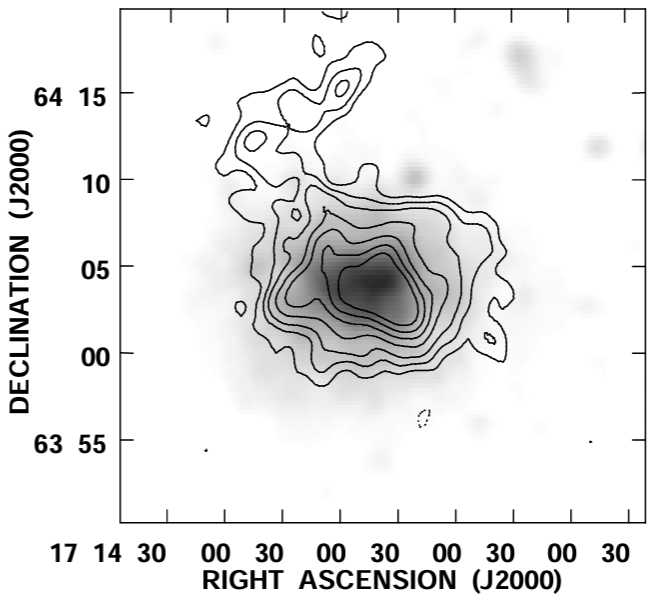}
    \caption{VLA 1.4\,GHz image \citep{feretti97}}
\end{subfigure}
\hspace{0.01\textwidth}
\begin{subfigure}[t]{0.4\textwidth}
    \centering
    \includegraphics[width=\linewidth,trim=0 0 0 0,clip]{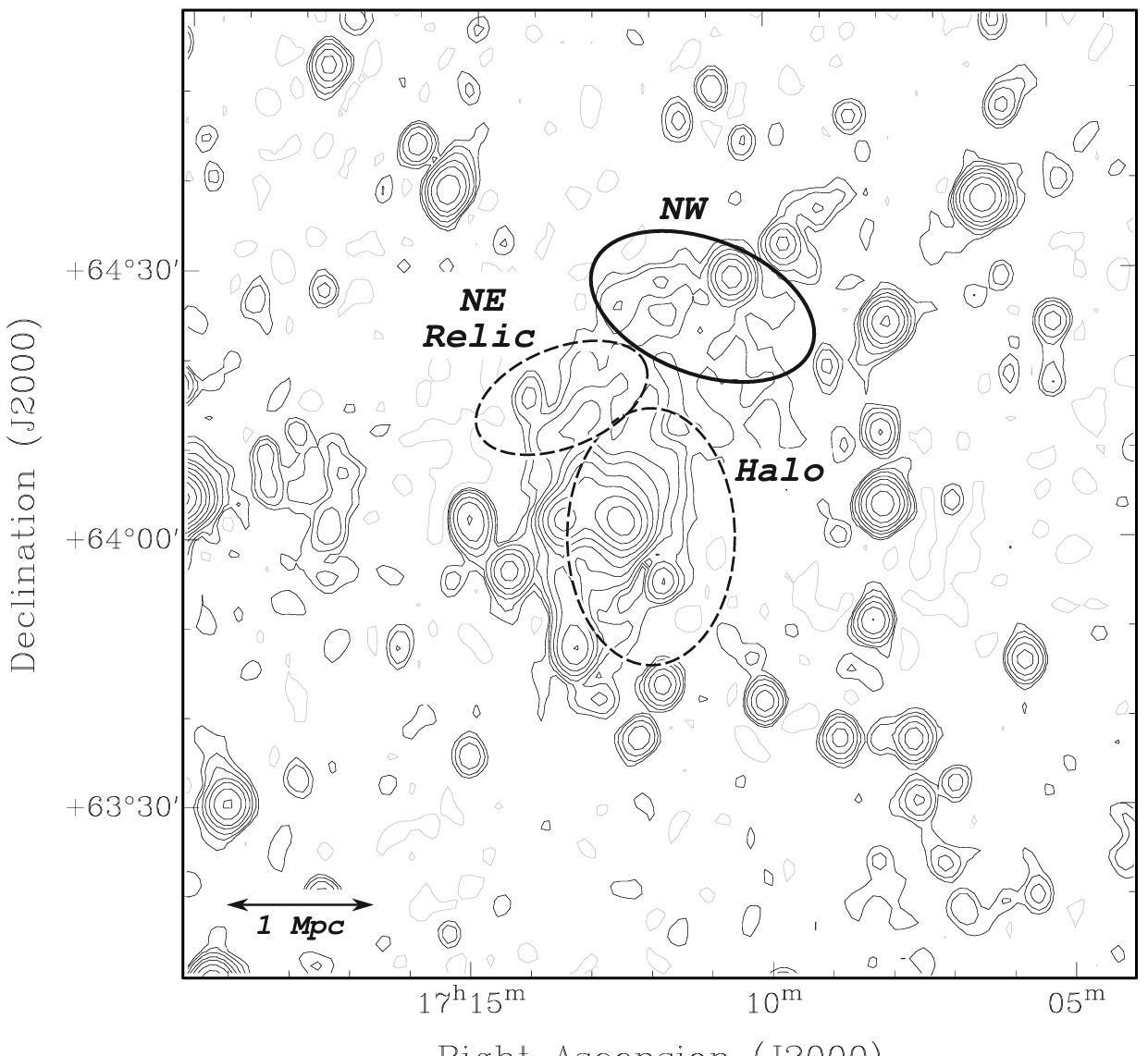}
    \caption{WSRT 350\,MHz image \citep{pizzo09}}
\end{subfigure}
\hspace{0.01\textwidth}
\begin{subfigure}[t]{0.5\textwidth}
    \centering
    \includegraphics[width=\linewidth,trim=0 0 0 0,clip]{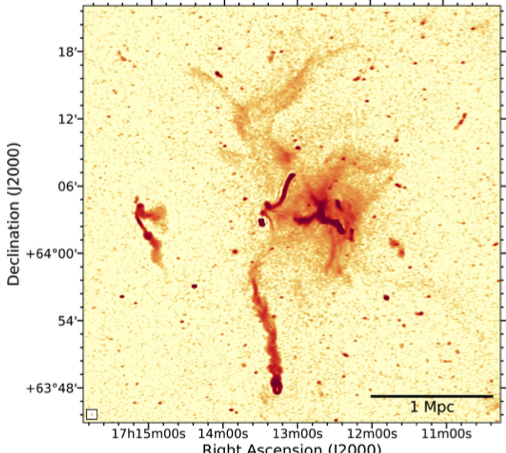}
    \caption{LOFAR 150\,MHz image \citep{Botteon20}}
\end{subfigure}
\caption{The radio halo in the galaxy cluster Abell~2255 observed with different radio facilities.}
\label{fig.a2255}
\end{figure}

\section{Advances from MeerKAT Studies}

The MeerKAT telescope has provided major advances in our understanding of diffuse, cluster-scale synchrotron emission. Owing to its excellent $uv$-coverage, high surface-brightness sensitivity, and $\sim$10 arcsec resolution at 1.28\,GHz, MeerKAT has revealed that radio halos are not smooth, amorphous structures but often contain filamentary substructures — narrow, elongated regions of enhanced synchrotron emission within the intracluster medium (ICM). Such features, typically a few to tens of kpc wide and extending up to several hundred kpc, have been detected in numerous systems, including Abell~3667, RXC~J1825.3+3026, and several MGCLS clusters \citep{Knowles2022,Riseley2022,Botteon2023,botteon24,Botteon25}. These suggest that magnetic-field amplification and turbulent 
enhanced synchrotron emissivity   
energy are distributed in a patchy, filamentary fashion
\citep{brunetti2014,Knowles2022}.

MeerKAT has also played a pivotal role in uncovering radio halos at intermediate and high redshifts  \citep[$z \gtrsim 1$;][]{Phuravhathu2025,Magolego2025,Sikhosana2025} and in establishing the population of ultra-steep spectrum radio halos \citep[e.g.][]{Knowles2022,rajpurohit21A2744,Magolego2025}.
These findings demonstrate that cluster-scale synchrotron emission was already established at earlier epochs than previously confirmed, implying efficient turbulence generation and magnetic-field amplification during structure formation.
In addition, the discovery of halos with very steep integrated spectra ($\alpha \gtrsim 1.7$--$2.0$) is consistent with the USSRH category predicted by turbulent re-acceleration models \citep{cassano2006,brunetti2014}. 
\citep{Phuravhathu2025,Magolego2025,Sikhosana2025}.

MeerKAT has also provided a substantial leap forward in the study of cluster magnetic fields and polarization. In Abell~2142, \citet{Pagliotta2025} derived a mean central magnetic field of $B_0 \simeq 9.5 \pm 1.0\,\mu$G and a magnetic-field power spectrum extending from $\sim7$ to $470$\,kpc. Using the ``Synchrotron Intensity Gradient'' (SIG) technique, \citet{Hu2023} mapped the orientation of magnetic fields in merging systems such as RXC~J1314.4$-$2515 and El~Gordo, finding alignment with the merger axis and small-scale turbulence. MeerKAT has also detected polarized filaments within halos and relics, tracing coherent magnetic features embedded in a turbulent ICM \citep{deGasperin2022}. 

\section{Advances from LOFAR and the LoTSS Surveys}
\label{Sect.LOFAR_advances}

The advent of the LOw Frequency ARray (LOFAR) has revolutionized the study of diffuse, cluster-scale radio emission by providing unprecedented sensitivity to low–surface-brightness, steep-spectrum sources in the 120--168\,MHz band. 
LOFAR fundamentally changed our view of diffuse emission in galaxy clusters by combining high sensitivity and angular resolution with excellent coverage of short baselines, enabling the recovery of both compact and extended structures with high fidelity. 

A striking example is provided by the galaxy cluster Abell~2255. As shown in Fig.~\ref{fig.a2255}, the well-known radio halo in this system was first imaged with the Very Large Array (VLA; \citealt{feretti97}) and later with the Westerbork Synthesis Radio Telescope (WSRT; \citealt{pizzo09}). 
The LOFAR image \citep[][Fig.~\ref{fig.a2255},panel~c]{Botteon20} reveals an unprecedented wealth of radio features, ranging from filamentary structures on scales of a few tens of kiloparsecs to diffuse cluster-scale emission. 
This comparison highlights LOFAR’s ability to unveil the full complexity of non-thermal phenomena in the intracluster medium, which had remained hidden in previous observations. Deep LOFAR imaging of the same cluster has unveiled pervasive synchrotron radio emission filling essentially the entire cluster volume and extending up to (or beyond) the virial radius, reaching up to $\sim 5$\,Mpc linear size or more, suggesting non‑thermal components filling even the cluster outskirts \citep{botteon22sci}. Additional observations revealed that some radio halos can be embedded within vast, ultra‑diffuse envelopes of synchrotron emission — so‑called megahalos \citep{cuciti22}. LOFAR also discovered synchrotron emission in the form of radio bridges connecting pairs of massive clusters \citep[\eg][]{govoni19,botteon18}. These evidences prove the presence of relativistic plasma on very large scale that could be better explored by SKA-Low \citep{Cuciti01.2026.SKA}.

Since 2015, LOFAR’s wide-area surveys, particularly the LOFAR Two-metre Sky Survey (LoTSS; \citealt{Shimwell2017,Shimwell2019,Shimwell2022}), have transformed our understanding of radio halos, relics, and mini-halos, extending detections to fainter and more distant systems than previously accessible and offering an unprecedented detailed view of these diffuse structures.
 
Using an SZ-selected sample of 309 Planck clusters, \citet{botteon2022} identified more than seventy radio halos, nearly doubling the number of known systems, and providing the first homogeneous statistical view of cluster-scale synchrotron emission at 150\,MHz. 
These observations confirmed that RHs are predominantly associated with merging and dynamically disturbed clusters, but also revealed new cases of diffuse emission in systems showing intermediate or relatively relaxed morphologies \citep{cassano+23}.
The resulting $P_{150\,\mathrm{MHz}}$--$M_{500}$ correlation (see Fig.\ref{Fig.Lrmin_z},~b)), spanning more than three orders of magnitude in radio power, shows a clear bimodality that separates merging from relaxed systems. Moreover, part of the observed scatter appears to originate from the different merger histories of clusters, reinforcing the scenario in which turbulence generated during mergers powers radio halos \citep{cuciti23}.

Building on these results, \citet{cassano+23} compared the LoTSS–DR2 statistics with Monte Carlo predictions based on turbulent re-acceleration theory.
They found an excellent agreement between observed and predicted distributions of RH number counts as a function of flux density and cluster redshift.
The occurrence of RH correlates strongly with merger activity and cluster mass, in line with turbulent re-acceleration models. 
Moreover, the fact that the fraction of clusters hosting RHs in LoTSS is higher than that measured in higher-frequency uGMRT observations suggests the presence of radio halos with very steep spectra ($\alpha \gtrsim 1.8$), detectable only at low frequencies, as predicted by these models. Several haloes with very steep spectra were detected at lower frequencies \citep[\eg][]{brunetti08nature,macario13, wilber18a1132, duchesne21arx2,bruno21,Rajpurohit2023}, thus supporting the idea that such a population might exist.
An important step to constrain the spectrum of RHs has been carried out by \citet{digennaro+21a,digennaro+21b}, who observed a small sample of massive clusters at high redshift ($z\geq0.6$) with LOFAR and then followed up the detected halos with the uGMRT. In line with models, about 50\% of these radio halos exhibit a very steep spectral index (\ie\ $\alpha \geq 1.5$ between 150–650 MHz) and are found among the less massive clusters in that sample.
  
\section{Detecting Radio Halos in Deep Radio Surveys}\label{sec:FRH}

\begin{figure}[t]
\centering
\begin{subfigure}[b]{0.45\textwidth}
    \centering
    \includegraphics[width=\linewidth, trim=0 100 0 0, clip]{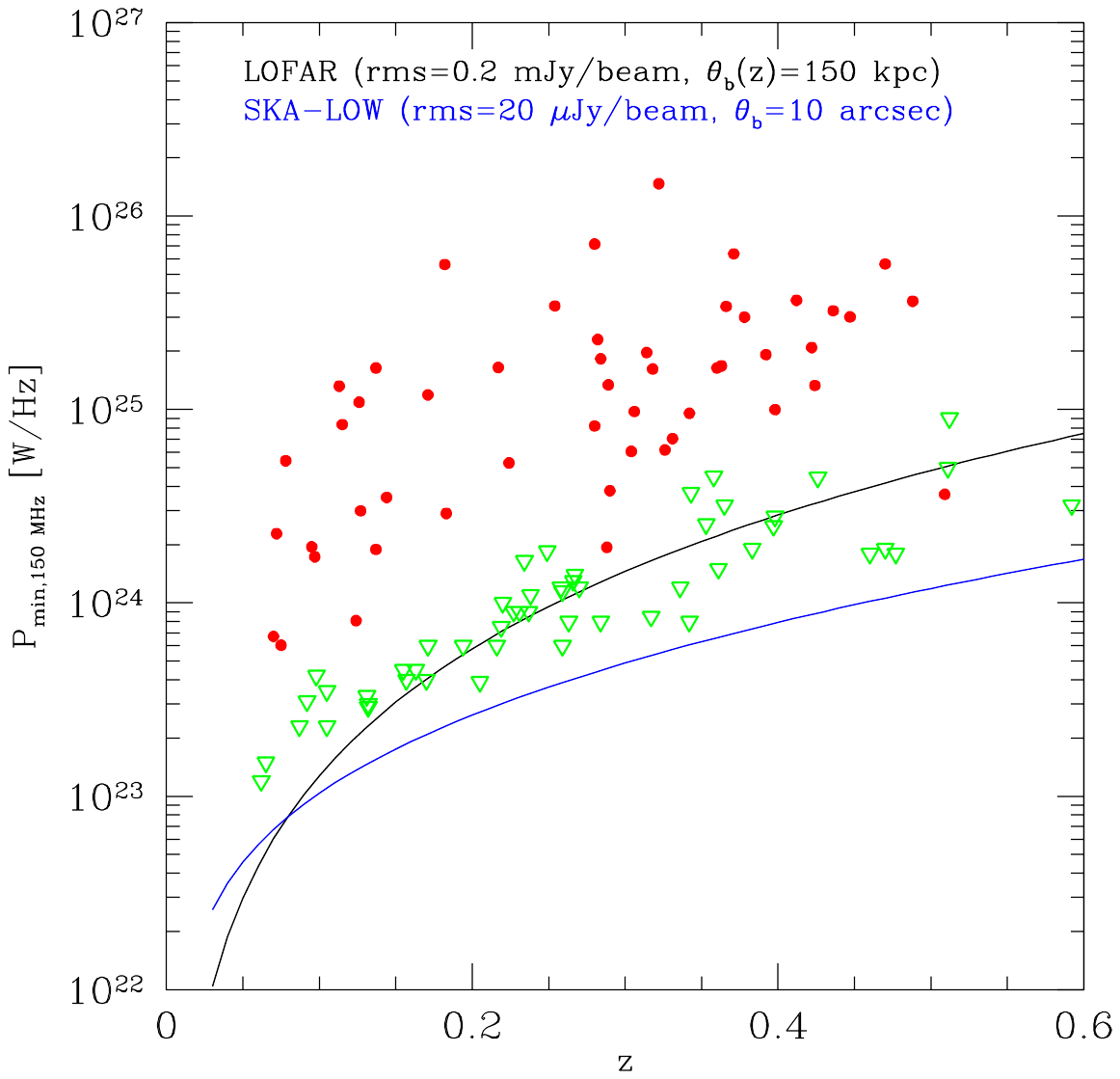}
    \caption{Radio power of halos (red points) and upper limits (green triangles) as a function of redshift from the LoTSS-DR2. 
The minimum radio power derived from Eq.~\ref{Eq.fmin} is shown as black line for LoTSS and as blu line for SKA-Low (parameters are as in figure panel) .}
\end{subfigure}
\hfill
\begin{subfigure}[b]{0.44\textwidth}
    \centering
    \includegraphics[width=\linewidth]{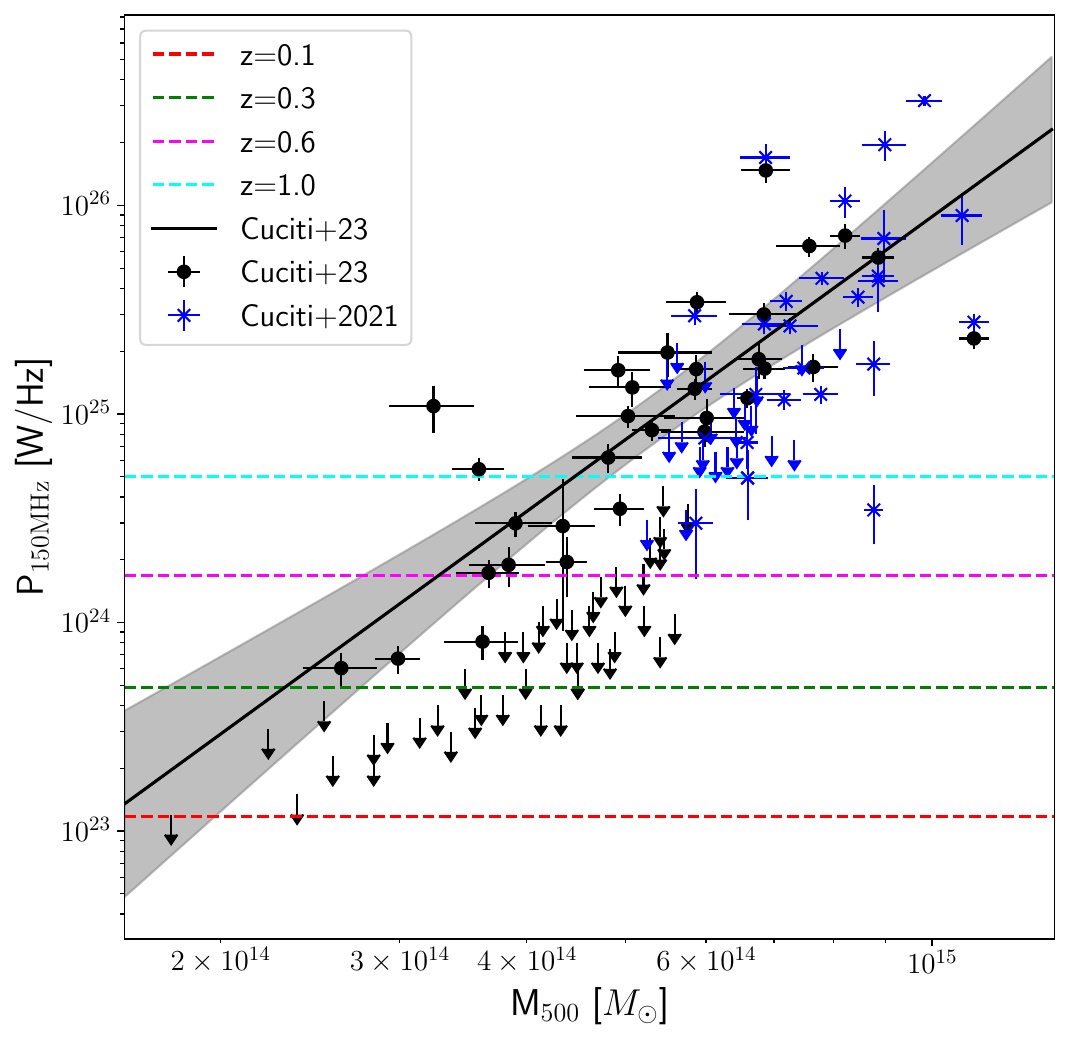}
    \caption{Radio power- mass relation: black points and arrows are RHs and upper limits from the LOFAR sample, blue stars and arrows are RHs and upper limits in \cite{cuciti21b}. The horizontal colored lines show the sensitivity of SKA-Low at different redshifts (see figure panel).}
\end{subfigure}
\caption{Radio power of halos and detection thresholds as a function of redshift.}
\label{Fig.Lrmin_z}
\end{figure}

In Fig.~\ref{Fig.Lrmin_z}, a) we show the measured radio powers (red points) and upper limits (green triangles) of cluster radio halos from the LoTSS-DR2 as a function of redshift.  
This diagram allows us to explore the minimum radio power of a halo that can be detected in a given survey, based on its sensitivity and angular resolution.  
We also report an analytic expression, previously used in several works \citep[e.g.][]{cassano10lofar,cassano12,cassano+23}, to estimate the minimum integrated flux density of a radio halo detectable in a survey under the assumption that the halo is considered detected when the integrated flux within $2\times \theta_e$ (with $\theta_e$ being the angular size corresponding to the $e$-folding radius $r_e$) reaches a given signal-to-noise ratio $\xi$.  
In this case, the total flux density within $2\times \theta_e$ can be expressed as $f_{\mathrm{min}}(<2\,\theta_e)\simeq0.75\,f_{\mathrm{min}}(<3\,\theta_e)\simeq\xi\,\sqrt{N_b}\times F_{\mathrm{rms}}$, where $N_{b}$ is the number of independent beams within $2\times \theta_e$.  
It follows that the minimum detectable flux is:

\begin{equation}
f_{\mathrm{min}}(<3\,\theta_e,z)\simeq4.44\times10^{-3}\, \xi\,
\left(\frac{\mathrm{F_{rms}}}{10 \mu\mathrm{Jy}}\right)
\left(\frac{10\,\mathrm{arcsec}}{\theta_b}\right)
\left(\frac{\theta_{e}(z)}{\mathrm{arcsec}}\right)\, \, [\mathrm{mJy}]
\label{Eq.fmin}
\end{equation}

\noindent where $F_{\mathrm{rms}}$ is the rms noise level (in $\mu$Jy\,beam$^{-1}$) and $\theta_b$ is the beam full width at half maximum (FWHM) in arcseconds.

The corresponding minimum radio power $P_{\mathrm{min}}(z)$ is reported in Fig.~\ref{Fig.Lrmin_z}, a) as a black line for the LoTSS-DR2 surveys and as a blue line for a SKA-Low survey. The LoTSS line has been computed assuming a survey sensitivity of $F_{\mathrm{rms}} = 200\,\mu$Jy\,beam$^{-1}$, and an effective beam size $\theta_b(z)$ that depends on redshift, corresponding to a fixed physical scale of 150\,kpc.  
We also assume $\theta_e$ corresponding to $r_e = 170$\,kpc, which is approximately the median $e$-folding radius derived for known radio halos. With this choice of parameters, Eq.~\ref{Eq.fmin} with $\xi = 5$ reproduces well the envelope traced by the observational upper limits as a function of redshift.
It is essential, particularly for low-frequency radio observations, to carefully assess the impact of confusion noise. Although LoTSS-DR2 is still well above the confusion limit, the much higher sensitivity of the SKA may be significantly affected.
For a one-hour SKA-Low observation, we estimate an rms noise of $F_{\rm rms} = 20~\mu\mathrm{Jy,beam^{-1}}$ with a $10$-arcsec beam (using Briggs = 0) at a central frequency of 150 MHz. Under these conditions,
a SKA-Low continuum survey would already be confusion-limited \citep{braun2014,braun2019}. Besides classical confusion noise, the identification of diffuse radio halos may also be affected by residual emission from compact and extended radio galaxies embedded within the cluster volume. This effect becomes increasingly important for low-surface-brightness halos and at high redshift, where the angular extent of the diffuse emission is reduced. Accurate source subtraction and multi-resolution imaging strategies will therefore be essential to fully exploit the sensitivity of SKA-Low.



The blue line in Fig.~\ref{Fig.Lrmin_z},a illustrates how the improved sensitivity of SKA-LOW relative to LOFAR would enable the potential detection of radio halos even in clusters where LOFAR observations could only provide upper limits. 
This relation therefore defines, at each redshift, the minimum radio power of a detectable halo within the sample, and can also be used in the $P_{150}$–$M_{500}$ plane to determine, at each redshift, the population of RHs and upper limits that SKA-Low will be able to probe (Fig.\ref{Fig.Lrmin_z},b.). It is clear that observations with SKA-Low at relatively low redshifts ($z \leq 0.3$) will unveil the population of RH in low massive clusters as $M_{500}\sim 2-3\times 10^{14}\,M_{\odot}$, while at very high redshift $z\gtrsim 1$ only quite massive clusters $M_{500}\gtrsim 4-5\times 10^{14}\,M_{\odot}$ would be accessible to SKA-Low.

The relation provided by Eq.\ref{Eq.fmin}
will be adopted in the following analysis to compare theoretical expectations with the observed number of radio halos (see Sect.~\ref{sec:NH} for details).

\section{Models of diffuse radio emission in galaxy clusters}

A detailed description of the theoretical model adopted in this work can be found in \citet[][]{cassano05} and \citet[][]{cassano2006}, with applications to predictions of RH statistics for future surveys with LOFAR, Apertif/WSRT, and ASKAP presented in \citet[][]{cassano10lofar,cassano12,cassano+23}.
In this Section, we summarise the key aspects of this theoretical framework and discuss its main implications for the statistical properties of RHs and their connection to the thermodynamical and dynamical state of galaxy clusters. The updated model is calibrated against the latest observational constraints from the LoTSS-DR2 and is employed here to derive predictions for forthcoming SKA-Low observations.
 
We consider a scenario in which giant RH forms during cluster–cluster mergers, as turbulence generated in the ICM re-accelerates relativistic electrons. The properties and cosmic evolution of RHs are modelled using a Monte Carlo approach that follows the hierarchical growth of dark matter halos through the extended Press–Schechter formalism \citep[][]{lacey93}. For each merger event in the simulated merger trees, the injection of turbulence is estimated within the volume swept by the infalling subcluster, bounded by ram-pressure stripping effects. The total turbulent energy is computed as a fraction $\eta_t$ ($\sim 0.1$–$0.3$) of the $P \,dV$ work done by the subcluster during its passage through the main cluster.
In these models the turbulent energy, acceleration rate and magnetic field per unit volume are considered constant within the RH volume \citep[i.e. homogeneous models,][]{cassano10lofar}.

A key prediction of turbulent re-acceleration models is that the synchrotron spectrum of RHs steepens above a characteristic frequency $\nu_s$, determined by the balance between particle acceleration and radiative losses \citep[e.g.][]{fujita03,cassano05}. The value of $\nu_s$ traces the energetics of the merger event and scales with the acceleration efficiency $\chi$ and the mean magnetic field strength $\langle B \rangle$ in the emitting volume as
$\nu_s \propto \langle B \rangle,\chi^2 / (\langle B \rangle^2 + B_{\rm cmb}^2)^2$,
where $B_{\rm cmb}=3.2(1+z)^2,\mu$G represents the equivalent magnetic field of the cosmic microwave background \citep[][]{cassano2006,cassano10lofar}. Monte Carlo simulations of cluster merger histories allow us to estimate $\chi$ based on the rate of turbulence generation and physical ICM conditions, and to derive the resulting dependence of $\nu_s$ on cluster mass, redshift, and merger parameters \citep[see e.g.][]{cassano05}.

The model includes three main parameters: the fraction of turbulent energy, $\eta_t$; the typical halo radius, $R_H$; and the average magnetic field strength, $\langle B \rangle$. Their effects on predicted RH statistics have been explored extensively in previous studies \citep[e.g.][]{cassano2006,cassano08revised}. Here, we adopt a reference set of parameters — $\langle B \rangle = 2\, \mu$G (consistent with \citealt{bonafede10}), $\eta_t = 0.2$, and $R_H \simeq 400$ kpc — which has been shown to reproduce the observed RH statistics at both low (LOFAR) and high (uGMRT) frequencies \citep[][]{cassano+19,botteon21ant,digennaro+21b}.
Most notably, this model successfully reproduces the statistical properties of RHs observed in the LoTSS-DR2 survey, including their redshift and flux density distributions and their occurrence as a function of cluster mass \citep[][]{cassano+23}. Furthermore, based on our previous works on the RH statistics at 1.4 GHz \citep[\eg][]{cassano06} and on the comparison between LOFAR and uGMRT RH statistics for high-z clusters \citep[\eg][]{digennaro+21a} we expect the general conclusions presented here to remain robust against reasonable variations of the adopted parameters.

\subsection{Occurrence of radio halos}

In the framework of the adopted scenario, the population of RHs is expected to consist of a diverse mixture of sources with different spectral properties. 
More massive (and hence hotter) clusters tend to host halos with flatter spectra, whereas less massive systems are predicted to produce steeper-spectrum halos.

To estimate the occurrence of RHs in a survey at a given observing frequency, $\nu_o$, we assume that only halos with a spectral steepening frequency $\nu_s \geq \nu_o$ are detectable. 
Energetic arguments suggest that halos with $\nu_s \geq 1$\,GHz are associated with the most powerful merger events in the Universe, since only such extreme interactions can inject sufficient turbulence on megaparsec scales to sustain the acceleration of relativistic electrons radiating above 1\,GHz \citep[][]{cassano05}. 

Within this framework, the fraction of clusters hosting RHs increases with cluster mass, as more massive systems experience stronger turbulent motions \citep[e.g.][]{vazza06,hallman11} and are therefore more likely to generate a radio halo. 
This expectation is consistent with GHz-frequency surveys, which detect RHs predominantly in the most massive and dynamically disturbed clusters \citep[e.g.][]{cassano2013,cuciti15,cuciti21b}. Conversely, RHs with lower $\nu_s$—that is, halos characterized by ultra–steep–spectrum emission (USSRHs, $\alpha>1.8$ between 250–600\,MHz.) are expected to be more common. 
These halos can originate from less energetic phenomena, such as major mergers between lower-mass systems or minor mergers involving massive clusters, both of which occur more frequently throughout cosmic time.

\begin{figure}[t]
\label{Fig.fRH}
\centering
\begin{subfigure}[t]{0.3\textwidth} 
    \centering
    \includegraphics[width=\linewidth,trim=0 0 0 0,clip]{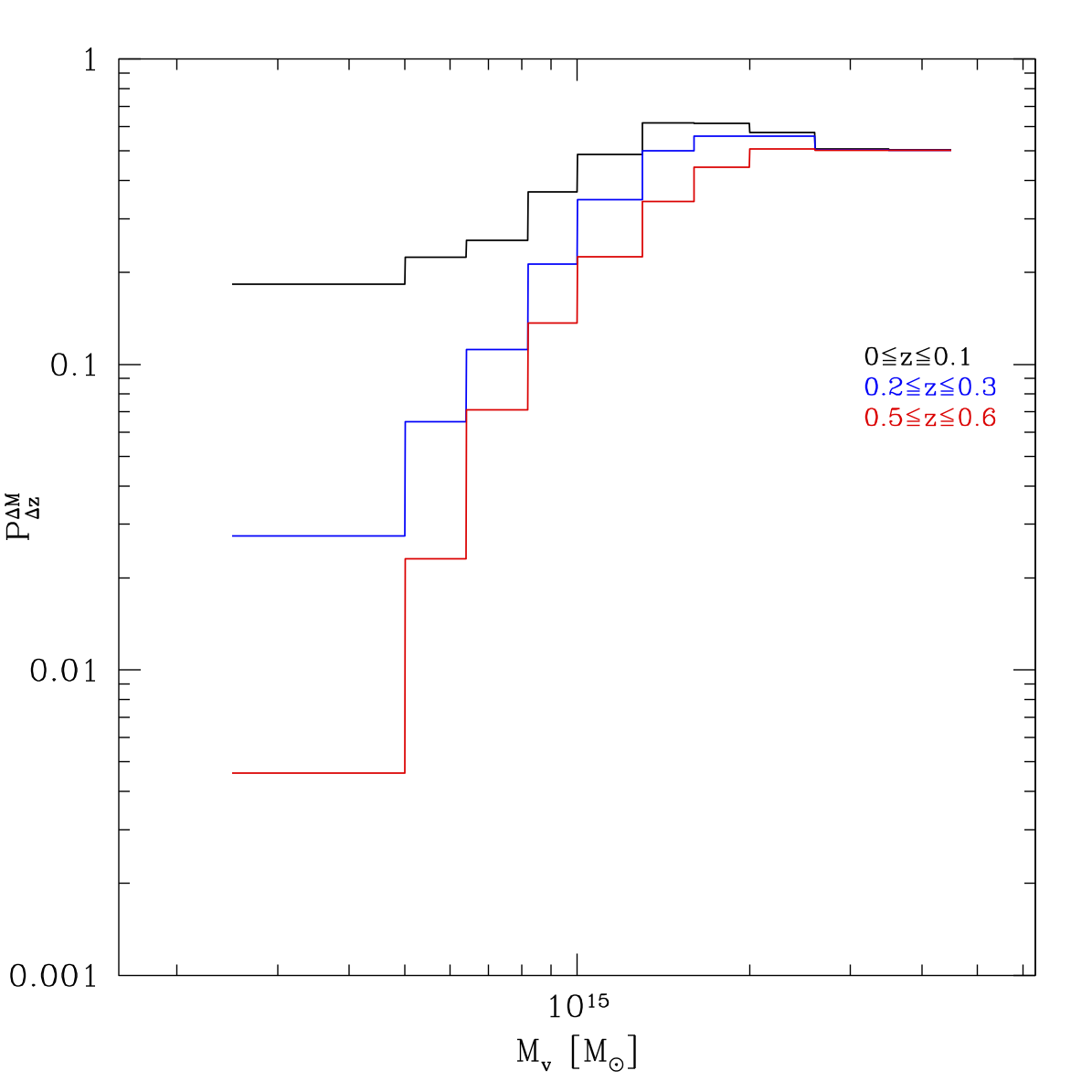}
    \caption{Expected fraction of clusters with RHs with $\nu_s \geq$ 150 MHz as a function of the cluster mass.}
\end{subfigure}
\hspace{0.01\textwidth}
\begin{subfigure}[t]{0.3\textwidth}
    \centering
    \includegraphics[width=\linewidth,trim=0 0 0 0,clip]{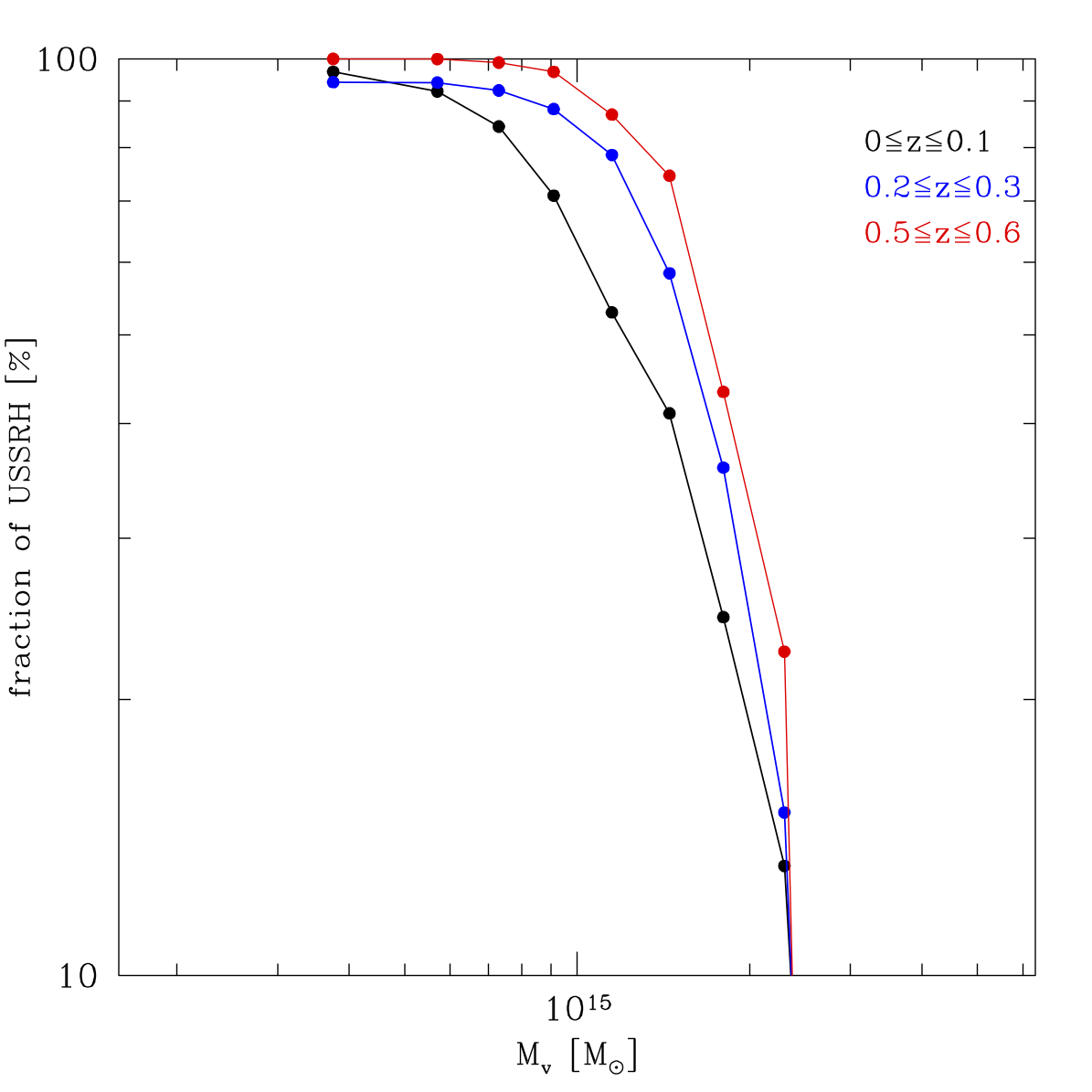}
    \caption{Expected fraction of RH with very steep radio spectra ($\nu_s<600$ MHz) as a function of the cluster mass.}
\end{subfigure}
\hspace{0.01\textwidth}
\begin{subfigure}[t]{0.3\textwidth}
    \centering
    \includegraphics[width=\linewidth,trim=0 0 0 0,clip]{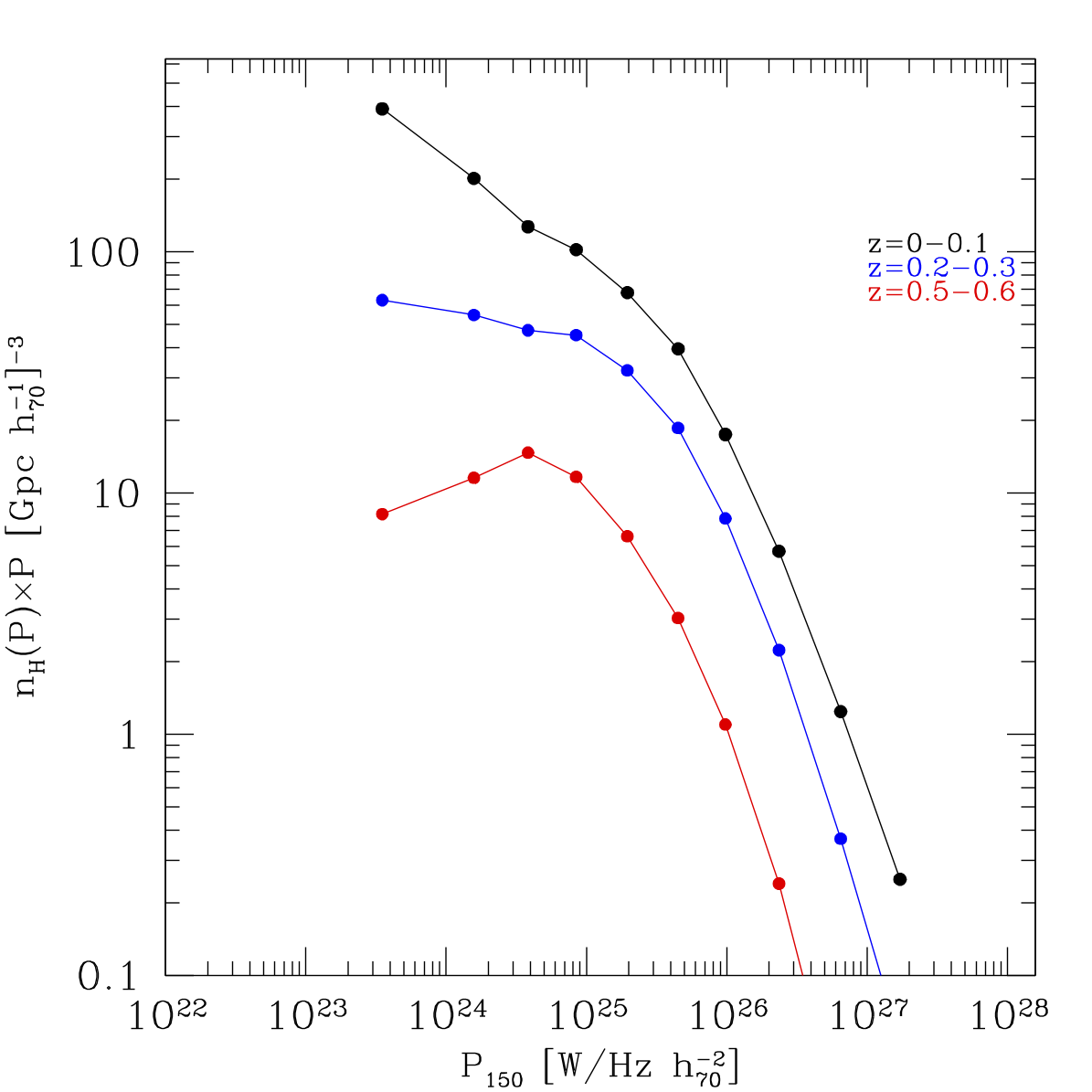}
    \caption{RH luminosity function at $\nu_o$=150 MHz. In all panels, the lines refer to z=$0-0.1$ (black line), $0.2-0.3$ (blue line), and $0.5-0.6$ (red line).}
\end{subfigure}
\caption{Model expectations at 150 MHz from \citep{cassano+23}.}
\label{Fig.fRH}
\end{figure}

\noindent \citet[][]{cassano12} showed that the fraction of clusters hosting RHs increases toward lower observing frequencies, and that the amplitude of this increase depends on the mass and redshift of the parent clusters—being more pronounced for less massive and higher-redshift systems.

In Fig.~\ref{Fig.fRH},a) we show the expected fraction of clusters hosting RHs with $\nu_s \geq 150$ MHz (black solid line) as a function of virial mass and redshift (see legend and caption), assuming the {\it reference} model parameters $\langle B \rangle = 2,\mu$G and $\eta_t = 0.2$. At all redshifts, this fraction increases with cluster mass. Fig.~\ref{Fig.fRH},b) shows the fraction of these RHs that are predicted to have ultra-steep spectra ($150 < \nu_s < 600$ MHz) under homogeneous models. The fraction of very steep-spectrum halos decreases with increasing cluster mass and varies with redshift.
For instance, at $z \simeq 0.05$, $\sim$80–90\% of RHs in clusters with $M_v \sim 8\times10^{14}~{\rm M_\odot}$ ($M_{500} \sim 4\times10^{14}~{\rm M_\odot}$) are expected to be very steep, dropping to $\sim$40\% for clusters with $M_v \sim 1.4\times10^{15},M_\odot$ ($M_{500} \sim 7\times10^{14}~{\rm M_\odot}$). At higher redshifts, the predicted fraction of USSRH rises for clusters of similar mass, reflecting stronger inverse Compton losses and the evolving merger rate.


\subsection{The radio halo luminosity function}
\label{Sect.RHLF}

The luminosity functions of radio halos (RHLFs) with $\nu_s\geq \nu_0$ 
(\ie\, the expected number of halos per comoving volume and radio power ``observable'' at
frequency $\nu_0$) 
can be estimated by :

\begin{equation}
{dN_{H}(z)\over{dV\,dP(\nu_0)}}=
{dN_{H}(z)\over{dM\,dV}}\bigg/ {dP(\nu_0)\over dM}\,,
\label{RHLF}
\end{equation}

\noindent
where $dN_{H}(z)/dM\,dV$ is the theoretical mass function of radio 
halos with $\nu_s \geq \nu_0$, that is obtained by combining Monte-Carlo
calculations of the fraction of clusters with RHs and the Press \& Schechter (PS) mass function 
of clusters \citep[\eg\,][]{cassano2006}.
We estimate $dP(\nu_0)/dM$ from the radio power-mass correlation which is given by \citep{cuciti23}:


\begin{equation}
\mathrm{log}\left(\frac{P_{150 \rm{MHz}}}{10^{24.5}\mathrm{W/Hz}}\right)=B~\mathrm{log}\left(\frac{M_{500}}{10^{14.9}\,M_\odot}\right)+A
\label{Eq:PM}
\end{equation}
with $A=1.1\pm0.1$ and $B=3.59\pm0.48$, which has a measured scatter $\sigma_{raw}\sim0.4$ \citep[see][]{cuciti23}.

The RHLFs at three different redshifts are reported in Fig.~\ref{Fig.fRH}, c). As already discussed in previous papers, the shape of the RHLF flattens at low radio powers because of the expected decrease in the efficiency of particle acceleration in the case of less massive clusters. However, the flattening at low power is less relevant than that expected considering observations at higher frequency because of the presence of radio halos with low values of $\nu_s$ contributing to the low power end of the luminosity function. Finally, we note that the normalisation of the RHLFs decreases with increasing redshift due to the evolution with $z$ of both the cluster mass function and the fraction of galaxy clusters with RHs \citep[Fig.~\ref{Fig.fRH}, a); see also][]{cassano2006}. 

The predictions presented in this work rely on the observed
$P_{150\,\mathrm{MHz}}$--$M_{500}$ scaling relation, which is characterized by uncertainties in both slope and normalization, as well as by an intrinsic scatter of $\sigma_{\rm raw}\sim0.4$ dex (Cuciti et al. 2023). Variations within the currently allowed range of these parameters mainly affect the predicted number of radio halos close to the survey detection threshold.
However, previous studies based on the same Monte Carlo framework have shown that the main trends discussed here remain robust against reasonable variations of the adopted scaling relation and model parameters. In particular, the increase of radio-halo occurrence with cluster mass, the expected presence of a large population of ultra-steep-spectrum halos, and the substantial gain in discovery space provided by SKA-Low are preserved
within the current observational uncertainties.

\section{Expected number of RH in SKA-Low}
\label{sec:NH}


In this Section, we present model predictions for the number of RHs detectable by SKA-LOW in AA4 configuration. As a reference, we adopt the same set of model parameters used in previous works ($\langle B \rangle=2\, \mu$G, $\eta_t=0.2$, $R_H\simeq 400$ kpc; Sect.\ref{sec:FRH}).

For SKA-LOW, the effective sensitivity and resolution determine the minimum flux $f_{min}(z)$ that a RH can have to be detected. The number of RHs within a redshift interval $\Delta z = z_2-z_1$ and with flux above $f_{min}(z)$ can be calculated by integrating the RHLF (Eq.~\ref{RHLF}):

\begin{equation}
N_{H}^{\Delta_z}=\int_{z=z_1}^{z=z_2}dz' \frac{dV}{dz'}
\int_{P_{min}(f_{min}^{*},z')} \frac{dN_H(P(\nu_o),z')}{dP(\nu_o)\,dV} dP(\nu_o)
\label{Eq.RHNC_SKA}
\end{equation}

The determination of $f_{min}(z)$ is now based on the expected sensitivity and resolution of SKA-Low \citep[e.g., $F_{rms}\sim 20\,\mu$Jy/beam and $\theta_b\sim10''$;][]{braun2019}. Given this sensitivity, the minimum radio power of detectable halos is derived and shown in Fig.\ref{Fig.Lrmin_z}, a) and Fig.\ref{fig:massz} as a function of redshift.
This provides a prediction of the RH population that would be observable in SKA-Low surveys over a given sky area.
The predicted cumulative number of RHs ($N_H(<z)$) and their distribution per redshift bin ($N_H(z,\Delta z)$) can be obtained by applying Eq.~\ref{Eq.RHNC_SKA} across the redshift range of interest. 



We estimate that SKA1-Low could be able to detect up to $\sim2600$ out to $z\sim0.6$ in the southern hemisphere. 
For comparison, the maximum number of halos detectable by LoTSS would be $\sim 1800$, in this case in the northern hemisphere. SKA-Low promises a tangible gain in the RH detection with respect to its precursors and pathfinders. 

\section{Radio halos at high-z}
\subsection{The lesson of LOFAR and MeerKAT}
The detection of radio halos with redshift greater than 0.6 \citep{cassano+19,digennaro+21a,digennaro+25a} is one of the main achievements of LOFAR 150 MHz observations. Despite the increase with $z$ of the Inverse Compton losses of the radio emitting electrons ($dE/dt \propto(1+z)^4$), the LOFAR observations of a small sample of 19 {\it Planck} clusters in the redshift range $0.6\leq z \leq0.9$ revealed an occurrence rate of about 50\%, in the mass range of $\rm 4-8\times10^{14}~M_\odot$ \citep{digennaro+21a}. This fraction decreases to $\sim9\%$ \citep{digennaro+25a} for a larger sample of clusters from the Massive and Distant Clusters of WISE Survey \citep[MaDCoWS;][]{gonzalez+19} in LoTSS-DR2, with lower masses ($M=1-6\times10^{14}~{\rm M_\odot}$) and higher redshifts ($0.78\leq z \leq 1.53$). 
Despite the bias in cluster selection, these results, along with serendipitous MeerKAT detections \citep{Phuravhathu2025,Magolego2025,Sikhosana2025}, indicate that diffuse Mpc-scale radio emission may already be present at earlier epochs. Moreover, and more importantly, these observations suggest $\rm\mu G$ magnetic field levels also at $z\sim0.7-1$, meaning a fast amplification of magnetic field seeds when the Universe was only 5--7 Gyr old if we assume radio halos are produced by turbulent re-acceleration mechanisms \citep{cassano05}. 
Consistent with this scenario, high-frequency follow-up observations with uGMRT Bands 3 and 4 of a sub-sample of {\it Planck} clusters hosting diffuse radio emission \citep{digennaro+21b} show that clusters in the ``low''-mass bin (i.e., $M_{500} < 5\times10^{14}~{\rm M_\odot}$) exhibit ultra-steep integrated spectra ($\alpha > 1.5$), whereas those in the ``high''-mass bin (i.e., $M_{500} > 5\times10^{14}~{\rm M_\odot}$) display flatter spectra ($\alpha \sim 1-1.3$).
This result is naturally limited by small-number statistics: the LOFAR–uGMRT follow-up included only nine clusters, of which only five were detected up to 650 MHz. 
Nevertheless, theoretical models designed to reproduce the radio halo statistics in nearby clusters \citep{cassano+23} are in a good agreement with these observations.

An important factor to take into account in the study of diffuse radio emission at high redshift is the proper subtraction of the radio emission associated with galaxies. For compact sources (i.e., those with sizes equal to the observing beam), this can be taken into account with a proper {\it uv}-cut. However, at extremely high redshift, the size of the radio halo can be of a few beams, making the subtraction from the {\it uv} plane difficult. For extended sources (i.e., extended radio galaxies and radio phoenices), a proper {\it uv}-subtraction is impossible, as the radio emission from the source could blend with that of the radio halo. For both cases, reaching high resolutions (sub-/arcsecond) is crucial. It has been shown that observations with the International LOFAR Telescope (ILT) at multiple resolutions can disentangle the different sources of radio emission \citep{hlavacek-larrondo+25}. Spectral index studies at high resolution can also help, as compact and extended radio galaxies are expected to have flatter spectral indices \citep[$\alpha\sim0.7-1.2$;][]{pinjarkar+25}.

\subsection{The expectations from SKA observations}
Despite the number of distant radio halos discovered with LOFAR and MeerKAT in the past few years, the current instrument sensitivities are limiting the detections to the most massive (and radio powerful) clusters (see black line in Fig. \ref{Fig.Lrmin_z}, a)). Given the current sensitivities of a standard 8-hour LOFAR observation of $\rm\sim200~\mu Jy\,beam^{-1}$ and resolution of $\sim20''$, \cite{digennaro+25a} has shown that radio halos at $z\sim0.6$ with radio power of $P_{\rm 150 MHz}>1.5\times10^{24}~{\rm W\,Hz^{-1}}$ can be detected. This minimum radio power increases by a factor of 3 and 10, for $z\sim1.0$ and $z\sim1.5$, respectively. The work-around could be increasing the observing time, but this would be very time-demanding as a 100-hour observation reaches a noise of $\rm\sim55~\mu Jy\,beam^{-1}$ \citep{tasse+21}. These long observations are therefore not feasible for large surveys of high-$z$ clusters, such as the Atacama Compact Telescope \citep[ACT;][]{calabrese+25,auguena26}, the South Pole Telescope \citep[SPT;][]{bocquet+19}, the SRG/eROSITA All-Sky Survey \citep[eRASS;][]{bulbul+24} and, in the future, {\it Euclid} \citep{euclidcoll+22}.

This picture will change with SKA-Low, where 1 hour of observations will be enough to reach a sensitivity of $\rm\sim20~\mu Jy\,beam^{-1}$ at a resolution of $\sim10''$, 
at the same frequency as LOFAR (i.e. 150 MHz, see blue line in Fig. \ref{Fig.Lrmin_z}). The main issue of these observations is that they reach the confusion limit, and consequently, separating diffuse radio emission and compact/extended radio galaxies will be challenging. 
The combination with SKA-Mid observations, which have higher resolutions and are not subject to confusion limits, will be crucial to extrapolating the flux density of the emission associated with the galaxies, assuming a standard spectral index.
The number counts presented in this work should therefore be regarded as an optimistic estimate based on thermal-noise sensitivity. In practice, confusion noise, imperfect subtraction of embedded radio galaxies, calibration artefacts, and surface-brightness limitations may reduce the completeness of radio-halo samples near the detection threshold. Nevertheless, the combination of SKA-Low with higher-resolution SKA-Mid observations is expected to mitigate these effects substantially, enabling robust identification of diffuse cluster-scale emission.


\section{The promise of SKA-Low AA4 cluster surveys}

\begin{figure}
\centering
\includegraphics[width=0.9\textwidth]{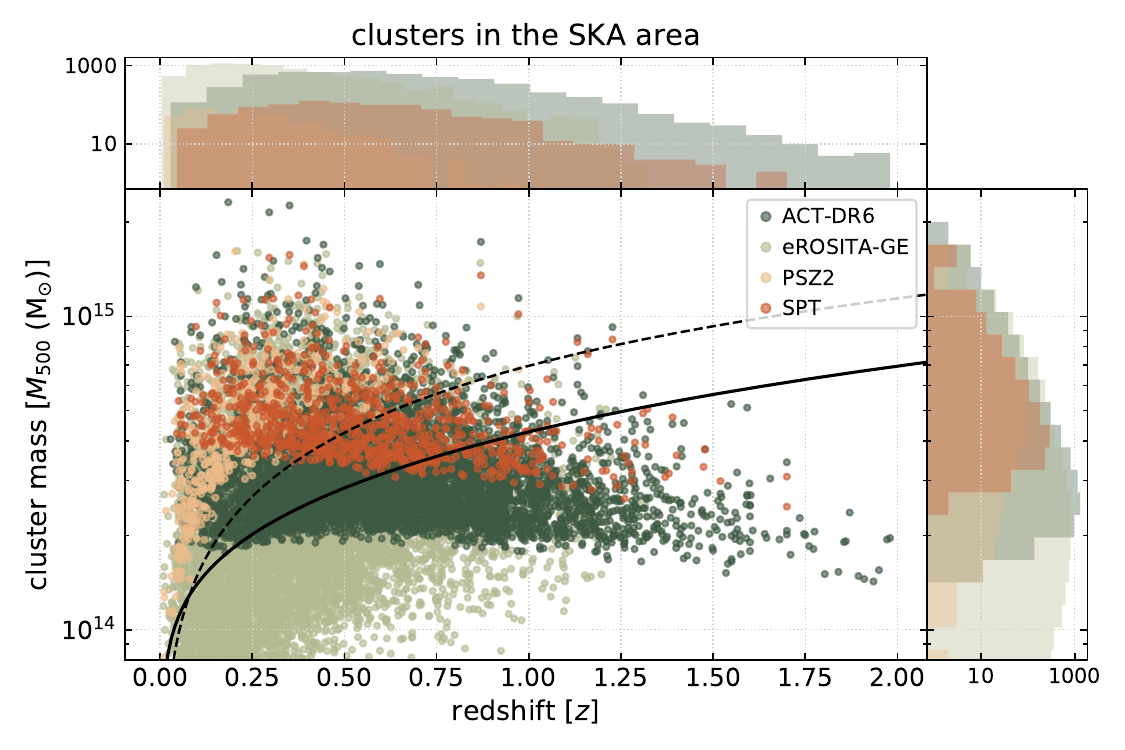}
\caption{Distribution of galaxy groups and clusters in the mass--redshift plane from current X-ray and SZ surveys (e.g.\ \textit{PSZ}, \textit{eROSITA}, \textit{SPT}, \textit{ACT}). Lines define for each redshift the minimum mass of clusters that will be efficiently probed by \textit{SKA-Low} in its AA4 configuration (solid line) and by LoTSS-DR2 (dashed line).}
\label{fig:massz}
\end{figure}

The SKA-Low telescope, particularly in its AA4 configuration, will provide an unprecedented view of diffuse non-thermal emission from galaxy clusters across the southern sky. This region includes the richest compilation of massive clusters currently known, thanks to recent and ongoing X-ray and Sunyaev--Zel'dovich (SZ) surveys such as \textit{eROSITA}, \textit{SPT}, and \textit{ACT} (Fig.~\ref{fig:massz}). These surveys span a wide region of the mass--redshift plane, extending to high redshift ($z \gtrsim 1$) and probing the most massive systems in the Universe. As shown in Fig.~\ref{fig:massz}, SKA-Low surveys in AA4 configuration will have access to a large fraction of these clusters, specifically: more than 4600 clusters at $z < 0.6$ and more than 700 at ($z \gtrsim 0.6$).

SKA-Low AA4 will enable the detection and characterization of radio halos in a mass range above $M_{500} \sim 10^{14}\,M_\odot$ and up to $z \sim 1$. 
The combination with SKA-Mid observations, which offer higher resolution and are not affected by confusion limits, will be crucial for subtracting the contribution of radio galaxies and AGN from the diffuse emission.

This capability will allow us to systematically probe the onset and evolution of non-thermal phenomena in clusters at earlier cosmic epochs, where current observations are extremely limited. In particular, SKA-Low surveys will: (i) provide the first statistically meaningful census of radio halos and ultra-steep-spectrum halos (USSRHs) in high-redshift systems $z>0.6$; (ii) test turbulent re-acceleration models 
in a cosmologically evolving environment; and (iii) reveal the connection between cluster mass assembly and the emergence of relativistic plasma and magnetic fields in the intracluster medium.

While the exact number of detectable radio halos depends on the adopted scaling relations and observational limitations, the predicted increase in discovery space relative to current LOFAR surveys remains substantial. Therefore, the qualitative conclusions of this work are not expected to be significantly affected by present uncertainties.

In summary, SKA-Low AA4 will complement and extend northern-sky surveys such as LoTSS, enabling a complete view of diffuse cluster emission across the sky and opening a new observational window on the formation and evolution of cosmic structure.

\bibliographystyle{abbrvnat-maxbibnames4}
\bibliography{chapter} 

@ARTICLE{Shimwell2017,
       author = {{Shimwell}, T.~W. and {R{\"o}ttgering}, H.~J.~A. and {Best}, P.~N. and {Williams}, W.~L. and {Dijkema}, T.~J. and {de Gasperin}, F. and {Hardcastle}, M.~J. and {Heald}, G.~H. and {Hoang}, D.~N. and {Horneffer}, A. and {Intema}, H. and {Mahony}, E.~K. and {Mandal}, S. and {Mechev}, A.~P. and {Morabito}, L. and {Oonk}, J.~B.~R. and {Rafferty}, D. and {Retana-Montenegro}, E. and {Sabater}, J. and {Tasse}, C. and {van Weeren}, R.~J. and {Br{\"u}ggen}, M. and {Brunetti}, G. and {Chy{\.z}y}, K.~T. and {Conway}, J.~E. and {Haverkorn}, M. and {Jackson}, N. and {Jarvis}, M.~J. and {McKean}, J.~P. and {Miley}, G.~K. and {Morganti}, R. and {White}, G.~J. and {Wise}, M.~W. and {van Bemmel}, I.~M. and {Beck}, R. and {Brienza}, M. and {Bonafede}, A. and {Calistro Rivera}, G. and {Cassano}, R. and {Clarke}, A.~O. and {Cseh}, D. and {Deller}, A. and {Drabent}, A. and {van Driel}, W. and {Engels}, D. and {Falcke}, H. and {Ferrari}, C. and {Fr{\"o}hlich}, S. and {Garrett}, M.~A. and {Harwood}, J.~J. and {Heesen}, V. and {Hoeft}, M. and {Horellou}, C. and {Israel}, F.~P. and {Kapi{\'n}ska}, A.~D. and {Kunert-Bajraszewska}, M. and {McKay}, D.~J. and {Mohan}, N.~R. and {Orr{\'u}}, E. and {Pizzo}, R.~F. and {Prandoni}, I. and {Schwarz}, D.~J. and {Shulevski}, A. and {Sipior}, M. and {Smith}, D.~J.~B. and {Sridhar}, S.~S. and {Steinmetz}, M. and {Stroe}, A. and {Varenius}, E. and {van der Werf}, P.~P. and {Zensus}, J.~A. and {Zwart}, J.~T.~L.},
        title = "{The LOFAR Two-metre Sky Survey. I. Survey description and preliminary data release}",
      journal = {\aap},
         year = 2017,
        month = feb,
       volume = {598},
          eid = {A104},
        pages = {A104},
          doi = {10.1051/0004-6361/201629313},
archivePrefix = {arXiv},
       eprint = {1611.02700},
 primaryClass = {astro-ph.IM},
       adsurl = {https://ui.adsabs.harvard.edu/abs/2017A&A...598A.104S}
}

@incollection{Cuciti01.2026.SKA, author = {Virginia Cuciti and author2 and author3 and author4 and author5},title = {},year = {2026},publisher = {},note = {arXiv search: Report number AASKAII/Cuciti01},booktitle = {Advancing Astrophysics with the SKA -- II (AASKAII)}}

@ARTICLE{auguena26,
       author = {{Aguena}, M. and {Aiola}, S. and {Allam}, S. and {Andrade-Oliveira}, F. and {Bacon}, D. and {Bahcall}, N. and {Battaglia}, N. and {Battistelli}, E.~S. and {Bocquet}, S. and {Bolliet}, B. and {Bond}, J.~R. and {Brooks}, D. and {Calabrese}, E. and {Carretero}, J. and {Choi}, S.~K. and {da Costa}, L.~N. and {Costanzi}, M. and {Coulton}, W. and {Davis}, T.~M. and {Desai}, S. and {Devlin}, M.~J. and {Dicker}, S. and {Doel}, P. and {Duivenvoorden}, A.~J. and {Dunkley}, J. and {Ferraro}, S. and {Flaugher}, B. and {Frieman}, J. and {Gallardo}, P.~A. and {Gatti}, M. and {Gaztanaga}, E. and {Gill}, A.~S. and {Golec}, J.~E. and {Gruen}, D. and {Gruendl}, R.~A. and {Halpern}, M. and {Hasselfield}, M. and {Hill}, J.~C. and {Hilton}, M. and {Hincks}, A.~D. and {Hinton}, S.~R. and {Hollowood}, D.~L. and {Honscheid}, K. and {Hubmayr}, J. and {Huffenberger}, K.~M. and {Hughes}, J.~P. and {James}, D.~J. and {Klein}, M. and {Knowles}, K. and {Koopman}, B.~J. and {Kosowsky}, A. and {Lahav}, O. and {Lee}, E. and {Lin}, Y. and {Lokken}, M. and {Madhavacheril}, M.~S. and {Malag{\'o}n}, A.~A. Plazas and {Marrewijk}, J. v. and {Marshall}, J.~L. and {McMahon}, J. and {Mena-Fern{\'a}ndez}, J. and {Miquel}, R. and {Miyatake}, H. and {Mohr}, J.~J. and {Moodley}, K. and {Mroczkowski}, T. and {Naess}, S. and {Nati}, F. and {Nicola}, A. and {Niemack}, M.~D. and {Ogando}, R.~L.~C. and {Oguri}, M. and {Orlowski-Scherer}, J. and {Page}, L.~A. and {Partridge}, B. and {da Silva Pereira}, M.~E. and {Porredon}, A. and {Qu}, F.~J. and {Ragavan}, D.~C. and {Guachalla}, B. Ried and {Romer}, A.~K. and {Rosell}, A. Carnero and {Rykoff}, E.~S. and {Samuroff}, S. and {Sanchez}, E. and {Sevilla-Noarbe}, I. and {Sierra}, C. and {Sif{\'o}n}, C. and {Smith}, M. and {Staggs}, S.~T. and {Suchyta}, E. and {Swanson}, M.~E.~C. and {Tucker}, D.~L. and {Vargas}, C. and {Vavagiakis}, E.~M. and {De Vicente}, J. and {Weaverdyck}, N. and {Weller}, J. and {Wollack}, E.~J. and {Zubeldia}, I.},
        title = "{The Atacama Cosmology Telescope: DR6 Sunyaev-Zel'dovich Selected Galaxy Clusters Catalog}",
      journal = {The Open Journal of Astrophysics},
         year = 2026,
        month = jan,
       volume = {9},
        pages = {55863},
          doi = {10.33232/001c.155863},
archivePrefix = {arXiv},
       eprint = {2507.21459},
 primaryClass = {astro-ph.CO},
       adsurl = {https://ui.adsabs.harvard.edu/abs/2026OJAp....955863A}
}

@article{macario13,
       author = {{Macario}, G. and {Venturi}, T. and {Intema}, H.~T. and {Dallacasa}, D. and {Brunetti}, G. and {Cassano}, R. and {Giacintucci}, S. and {Ferrari}, C. and {Ishwara-Chandra}, C.~H. and {Athreya}, R.},
        title = "{153 MHz GMRT follow-up of steep-spectrum diffuse emission in galaxy clusters}",
      journal = {\aap},
         year = 2013,
        month = mar,
       volume = {551},
          eid = {A141},
        pages = {A141},
          doi = {10.1051/0004-6361/201220667},
archivePrefix = {arXiv},
       eprint = {1302.0648},
 primaryClass = {astro-ph.CO},
       adsurl = {https://ui.adsabs.harvard.edu/abs/2013A&A...551A.141M}
}

@article{wilber18a1132,
       author = {{Wilber}, A. and {Br{\"u}ggen}, M. and {Bonafede}, A. and {Savini}, F. and {Shimwell}, T. and {van Weeren}, R.~J. and {Rafferty}, D. and {Mechev}, A.~P. and {Intema}, H. and {Andrade-Santos}, F. and {Clarke}, A.~O. and {Mahony}, E.~K. and {Morganti}, R. and {Prandoni}, I. and {Brunetti}, G. and {R{\"o}ttgering}, H. and {Mandal}, S. and {de Gasperin}, F. and {Hoeft}, M.},
        title = "{LOFAR discovery of an ultra-steep radio halo and giant head-tail radio galaxy in Abell 1132}",
      journal = {\mnras},
         year = 2018,
        month = jan,
       volume = {473},
       number = {3},
        pages = {3536-3546},
          doi = {10.1093/mnras/stx2568},
archivePrefix = {arXiv},
       eprint = {1708.08928},
 primaryClass = {astro-ph.GA},
       adsurl = {https://ui.adsabs.harvard.edu/abs/2018MNRAS.473.3536W}
}

@article{duchesne21arx2,
       author = {{Duchesne}, S.~W. and {Johnston-Hollitt}, M. and {Bartalucci}, I.},
        title = "{Low-frequency integrated radio spectra of diffuse, steep-spectrum sources in galaxy clusters: palaeontology with the MWA and ASKAP}",
      journal = {\pasa},
         year = 2021,
        month = oct,
       volume = {38},
          eid = {e053},
        pages = {e053},
          doi = {10.1017/pasa.2021.45},
archivePrefix = {arXiv},
       eprint = {2106.12281},
 primaryClass = {astro-ph.CO},
       adsurl = {https://ui.adsabs.harvard.edu/abs/2021PASA...38...53D}
}

@ARTICLE{pizzo09,
       author = {{Pizzo}, R.~F. and {de Bruyn}, A.~G.},
        title = "{Radio spectral study of the cluster of galaxies Abell 2255}",
      journal = {\aap},
         year = 2009,
        month = nov,
       volume = {507},
       number = {2},
        pages = {639-659},
          doi = {10.1051/0004-6361/200912465},
archivePrefix = {arXiv},
       eprint = {0909.5198},
 primaryClass = {astro-ph.CO},
       adsurl = {https://ui.adsabs.harvard.edu/abs/2009A&A...507..639P}
}

@ARTICLE{feretti97,
       author = {{Feretti}, L. and {Boehringer}, H. and {Giovannini}, G. and {Neumann}, D.},
        title = "{The radio and X-ray properties of Abell 2255.}",
      journal = {\aap},
         year = 1997,
        month = jan,
       volume = {317},
        pages = {432-440},
          doi = {10.48550/arXiv.astro-ph/9607027},
archivePrefix = {arXiv},
       eprint = {astro-ph/9607027},
 primaryClass = {astro-ph},
       adsurl = {https://ui.adsabs.harvard.edu/abs/1997A&A...317..432F}
}

@ARTICLE{Shimwell2019,
       author = {{Shimwell}, T.~W. and {Tasse}, C. and {Hardcastle}, M.~J. and {Mechev}, A.~P. and {Williams}, W.~L. and {Best}, P.~N. and {R{\"o}ttgering}, H.~J.~A. and {Callingham}, J.~R. and {Dijkema}, T.~J. and {de Gasperin}, F. and {Hoang}, D.~N. and {Hugo}, B. and {Mirmont}, M. and {Oonk}, J.~B.~R. and {Prandoni}, I. and {Rafferty}, D. and {Sabater}, J. and {Smirnov}, O. and {van Weeren}, R.~J. and {White}, G.~J. and {Atemkeng}, M. and {Bester}, L. and {Bonnassieux}, E. and {Br{\"u}ggen}, M. and {Brunetti}, G. and {Chy{\.z}y}, K.~T. and {Cochrane}, R. and {Conway}, J.~E. and {Croston}, J.~H. and {Danezi}, A. and {Duncan}, K. and {Haverkorn}, M. and {Heald}, G.~H. and {Iacobelli}, M. and {Intema}, H.~T. and {Jackson}, N. and {Jamrozy}, M. and {Jarvis}, M.~J. and {Lakhoo}, R. and {Mevius}, M. and {Miley}, G.~K. and {Morabito}, L. and {Morganti}, R. and {Nisbet}, D. and {Orr{\'u}}, E. and {Perkins}, S. and {Pizzo}, R.~F. and {Schrijvers}, C. and {Smith}, D.~J.~B. and {Vermeulen}, R. and {Wise}, M.~W. and {Alegre}, L. and {Bacon}, D.~J. and {van Bemmel}, I.~M. and {Beswick}, R.~J. and {Bonafede}, A. and {Botteon}, A. and {Bourke}, S. and {Brienza}, M. and {Calistro Rivera}, G. and {Cassano}, R. and {Clarke}, A.~O. and {Conselice}, C.~J. and {Dettmar}, R.~J. and {Drabent}, A. and {Dumba}, C. and {Emig}, K.~L. and {En{\ss}lin}, T.~A. and {Ferrari}, C. and {Garrett}, M.~A. and {G{\'e}nova-Santos}, R.~T. and {Goyal}, A. and {G{\"u}rkan}, G. and {Hale}, C. and {Harwood}, J.~J. and {Heesen}, V. and {Hoeft}, M. and {Horellou}, C. and {Jackson}, C. and {Kokotanekov}, G. and {Kondapally}, R. and {Kunert-Bajraszewska}, M. and {Mahatma}, V. and {Mahony}, E.~K. and {Mandal}, S. and {McKean}, J.~P. and {Merloni}, A. and {Mingo}, B. and {Miskolczi}, A. and {Mooney}, S. and {Nikiel-Wroczy{\'n}ski}, B. and {O'Sullivan}, S.~P. and {Quinn}, J. and {Reich}, W. and {Roskowi{\'n}ski}, C. and {Rowlinson}, A. and {Savini}, F. and {Saxena}, A. and {Schwarz}, D.~J. and {Shulevski}, A. and {Sridhar}, S.~S. and {Stacey}, H.~R. and {Urquhart}, S. and {van der Wiel}, M.~H.~D. and {Varenius}, E. and {Webster}, B. and {Wilber}, A.},
        title = "{The LOFAR Two-metre Sky Survey. II. First data release}",
      journal = {\aap},
         year = 2019,
        month = feb,
       volume = {622},
          eid = {A1},
        pages = {A1},
          doi = {10.1051/0004-6361/201833559},
archivePrefix = {arXiv},
       eprint = {1811.07926},
 primaryClass = {astro-ph.GA},
       adsurl = {https://ui.adsabs.harvard.edu/abs/2019A&A...622A...1S}
}

@ARTICLE{Shimwell2022,
       author = {{Shimwell}, T.~W. and {Hardcastle}, M.~J. and {Tasse}, C. and {Best}, P.~N. and {R{\"o}ttgering}, H.~J.~A. and {Williams}, W.~L. and {Botteon}, A. and {Drabent}, A. and {Mechev}, A. and {Shulevski}, A. and {van Weeren}, R.~J. and {Bester}, L. and {Br{\"u}ggen}, M. and {Brunetti}, G. and {Callingham}, J.~R. and {Chy{\.z}y}, K.~T. and {Conway}, J.~E. and {Dijkema}, T.~J. and {Duncan}, K. and {de Gasperin}, F. and {Hale}, C.~L. and {Haverkorn}, M. and {Hugo}, B. and {Jackson}, N. and {Mevius}, M. and {Miley}, G.~K. and {Morabito}, L.~K. and {Morganti}, R. and {Offringa}, A. and {Oonk}, J.~B.~R. and {Rafferty}, D. and {Sabater}, J. and {Smith}, D.~J.~B. and {Schwarz}, D.~J. and {Smirnov}, O. and {O'Sullivan}, S.~P. and {Vedantham}, H. and {White}, G.~J. and {Albert}, J.~G. and {Alegre}, L. and {Asabere}, B. and {Bacon}, D.~J. and {Bonafede}, A. and {Bonnassieux}, E. and {Brienza}, M. and {Bilicki}, M. and {Bonato}, M. and {Calistro Rivera}, G. and {Cassano}, R. and {Cochrane}, R. and {Croston}, J.~H. and {Cuciti}, V. and {Dallacasa}, D. and {Danezi}, A. and {Dettmar}, R.~J. and {Di Gennaro}, G. and {Edler}, H.~W. and {En{\ss}lin}, T.~A. and {Emig}, K.~L. and {Franzen}, T.~M.~O. and {Garc{\'\i}a-Vergara}, C. and {Grange}, Y.~G. and {G{\"u}rkan}, G. and {Hajduk}, M. and {Heald}, G. and {Heesen}, V. and {Hoang}, D.~N. and {Hoeft}, M. and {Horellou}, C. and {Iacobelli}, M. and {Jamrozy}, M. and {Jeli{\'c}}, V. and {Kondapally}, R. and {Kukreti}, P. and {Kunert-Bajraszewska}, M. and {Magliocchetti}, M. and {Mahatma}, V. and {Ma{\l}ek}, K. and {Mandal}, S. and {Massaro}, F. and {Meyer-Zhao}, Z. and {Mingo}, B. and {Mostert}, R.~I.~J. and {Nair}, D.~G. and {Nakoneczny}, S.~J. and {Nikiel-Wroczy{\'n}ski}, B. and {Orr{\'u}}, E. and {Pajdosz-{\'S}mierciak}, U. and {Pasini}, T. and {Prandoni}, I. and {van Piggelen}, H.~E. and {Rajpurohit}, K. and {Retana-Montenegro}, E. and {Riseley}, C.~J. and {Rowlinson}, A. and {Saxena}, A. and {Schrijvers}, C. and {Sweijen}, F. and {Siewert}, T.~M. and {Timmerman}, R. and {Vaccari}, M. and {Vink}, J. and {West}, J.~L. and {Wo{\l}owska}, A. and {Zhang}, X. and {Zheng}, J.},
        title = "{The LOFAR Two-metre Sky Survey. V. Second data release}",
      journal = {\aap},
         year = 2022,
        month = mar,
       volume = {659},
          eid = {A1},
        pages = {A1},
          doi = {10.1051/0004-6361/202142484},
archivePrefix = {arXiv},
       eprint = {2202.11733},
 primaryClass = {astro-ph.GA},
       adsurl = {https://ui.adsabs.harvard.edu/abs/2022A&A...659A...1S}
}

@ARTICLE{botteon24,
       author = {{Botteon}, A. and {van Weeren}, R.~J. and {Eckert}, D. and {Gastaldello}, F. and {Markevitch}, M. and {Giacintucci}, S. and {Brunetti}, G. and {Kale}, R. and {Venturi}, T.},
        title = "{The prototypical major cluster merger Abell 754: I. Calibration of MeerKAT data and radio/X-ray spectral mapping of the cluster}",
      journal = {\aap},
         year = 2024,
        month = oct,
       volume = {690},
          eid = {A222},
        pages = {A222},
          doi = {10.1051/0004-6361/202451293},
archivePrefix = {arXiv},
       eprint = {2406.18983},
 primaryClass = {astro-ph.CO},
       adsurl = {https://ui.adsabs.harvard.edu/abs/2024A&A...690A.222B}
}

@ARTICLE{kim25,
       author = {{HyeongHan}, Kim and {Jee}, M. James and {Lee}, Wonki and {ZuHone}, John and {Zhuravleva}, Irina and {Kang}, Wooseok and {Hwang}, Ho Seong},
        title = "{Direct evidence of a major merger in the Perseus cluster}",
      journal = {Nature Astronomy},
         year = 2025,
        month = jun,
       volume = {9},
        pages = {925-931},
          doi = {10.1038/s41550-025-02530-w},
archivePrefix = {arXiv},
       eprint = {2405.00115},
 primaryClass = {astro-ph.GA},
       adsurl = {https://ui.adsabs.harvard.edu/abs/2025NatAs...9..925H}
}

@article{cassano2013,
       author = {{Cassano}, R. and {Ettori}, S. and {Brunetti}, G. and {Giacintucci}, S. and {Pratt}, G.~W. and {Venturi}, T. and {Kale}, R. and {Dolag}, K. and {Markevitch}, M.},
        title = "{Revisiting Scaling Relations for Giant Radio Halos in Galaxy Clusters}",
      journal = {\apj},
         year = 2013,
        month = nov,
       volume = {777},
       number = {2},
          eid = {141},
        pages = {141},
          doi = {10.1088/0004-637X/777/2/141},
archivePrefix = {arXiv},
       eprint = {1306.4379},
 primaryClass = {astro-ph.CO},
       adsurl = {https://ui.adsabs.harvard.edu/abs/2013ApJ...777..141C}
}

@article{santra2024deep,
       author = {{Santra}, R. and {Kale}, R. and {Giacintucci}, S. and {Markevitch}, M. and {De Luca}, F. and {Bourdin}, H. and {Venturi}, T. and {Dallacasa}, D. and {Cassano}, R. and {Brunetti}, G. and {Buch}, K.~D.},
        title = "{A Deep uGMRT View of the Ultra-steep Spectrum Radio Halo in A521}",
      journal = {\apj},
         year = 2024,
        month = feb,
       volume = {962},
       number = {1},
          eid = {40},
        pages = {40},
          doi = {10.3847/1538-4357/ad1190},
archivePrefix = {arXiv},
       eprint = {2311.09717},
 primaryClass = {astro-ph.CO},
       adsurl = {https://ui.adsabs.harvard.edu/abs/2024ApJ...962...40S}
}

@ARTICLE{dallacasa09,
       author = {{Dallacasa}, D. and {Brunetti}, G. and {Giacintucci}, S. and {Cassano}, R. and {Venturi}, T. and {Macario}, G. and {Kassim}, N.~E. and {Lane}, W. and {Setti}, G.},
        title = "{Deep 1.4 GHz Follow-up of the Steep Spectrum Radio Halo in A521}",
      journal = {\apj},
         year = 2009,
        month = jul,
       volume = {699},
       number = {2},
        pages = {1288-1292},
          doi = {10.1088/0004-637X/699/2/1288},
archivePrefix = {arXiv},
       eprint = {0905.0588},
 primaryClass = {astro-ph.CO},
       adsurl = {https://ui.adsabs.harvard.edu/abs/2009ApJ...699.1288D}
}

@article{bruno21,
       author = {{Bruno}, L. and {Rajpurohit}, K. and {Brunetti}, G. and {Gastaldello}, F. and {Botteon}, A. and {Ignesti}, A. and {Bonafede}, A. and {Dallacasa}, D. and {Cassano}, R. and {van Weeren}, R.~J. and {Cuciti}, V. and {Di Gennaro}, G. and {Shimwell}, T. and {Br{\"u}ggen}, M.},
        title = "{The LOFAR and JVLA view of the distant steep spectrum radio halo in MACS J1149.5+2223}",
      journal = {\aap},
         year = 2021,
        month = jun,
       volume = {650},
          eid = {A44},
        pages = {A44},
          doi = {10.1051/0004-6361/202039877},
archivePrefix = {arXiv},
       eprint = {2103.10110},
 primaryClass = {astro-ph.CO},
       adsurl = {https://ui.adsabs.harvard.edu/abs/2021A&A...650A..44B}
}

@ARTICLE{cuciti23,
       author = {{Cuciti}, V. and {Cassano}, R. and {Sereno}, M. and {Brunetti}, G. and {Botteon}, A. and {Shimwell}, T.~W. and {Bruno}, L. and {Gastaldello}, F. and {Rossetti}, M. and {Zhang}, X. and {Simionescu}, A. and {Br{\"u}ggen}, M. and {van Weeren}, R.~J. and {Jones}, A. and {Akamatsu}, H. and {Bonafede}, A. and {De Gasperin}, F. and {Di Gennaro}, G. and {Pasini}, T. and {R{\"o}ttgering}, H.~J.~A.},
        title = "{The Planck clusters in the LOFAR sky. V. LoTSS-DR2: Mass-radio halo power correlation at low frequency}",
      journal = {\aap},
         year = 2023,
        month = dec,
       volume = {680},
          eid = {A30},
        pages = {A30},
          doi = {10.1051/0004-6361/202346755},
archivePrefix = {arXiv},
       eprint = {2305.04564},
 primaryClass = {astro-ph.CO},
       adsurl = {https://ui.adsabs.harvard.edu/abs/2023A&A...680A..30C}
}

@ARTICLE{vanweeren24pers,
       author = {{van Weeren}, R.~J. and {Timmerman}, R. and {Vaidya}, V. and {Gendron-Marsolais}, M.-L. and {Botteon}, A. and {Roberts}, I.~D. and {Hlavacek-Larrondo}, J. and {Bonafede}, A. and {Br{\"u}ggen}, M. and {Brunetti}, G. and {Cassano}, R. and {Cuciti}, V. and {Edge}, A.~C. and {Gastaldello}, F. and {Groeneveld}, C. and {Shimwell}, T.~W.},
        title = "{LOFAR high-band antenna observations of the Perseus cluster: The discovery of a giant radio halo}",
      journal = {\aap},
         year = 2024,
        month = dec,
       volume = {692},
          eid = {A12},
        pages = {A12},
          doi = {10.1051/0004-6361/202451618},
archivePrefix = {arXiv},
       eprint = {2410.02863},
 primaryClass = {astro-ph.CO},
       adsurl = {https://ui.adsabs.harvard.edu/abs/2024A&A...692A..12V}
}

@ARTICLE{Botteon25,
       author = {{Botteon}, Andrea and {Balboni}, Marco and {Bartalucci}, Iacopo and {Gastaldello}, Fabio and {van Weeren}, Reinout J.},
        title = "{MeerKAT L-band observations of the Ophiuchus galaxy cluster: Detection of synchrotron threads and jellyfish galaxies}",
      journal = {\aap},
         year = 2025,
        month = jun,
       volume = {698},
          eid = {A55},
        pages = {A55},
          doi = {10.1051/0004-6361/202554695},
archivePrefix = {arXiv},
       eprint = {2504.16158},
 primaryClass = {astro-ph.GA},
       adsurl = {https://ui.adsabs.harvard.edu/abs/2025A&A...698A..55B}
}

@ARTICLE{Botteon20,
       author = {{Botteon}, A. and {Brunetti}, G. and {van Weeren}, R.~J. and {Shimwell}, T.~W. and {Pizzo}, R.~F. and {Cassano}, R. and {Iacobelli}, M. and {Gastaldello}, F. and {B{\^\i}rzan}, L. and {Bonafede}, A. and {Br{\"u}ggen}, M. and {Cuciti}, V. and {Dallacasa}, D. and {de Gasperin}, F. and {Di Gennaro}, G. and {Drabent}, A. and {Hardcastle}, M.~J. and {Hoeft}, M. and {Mandal}, S. and {R{\"o}ttgering}, H.~J.~A. and {Simionescu}, A.},
        title = "{The Beautiful Mess in Abell 2255}",
      journal = {\apj},
         year = 2020,
        month = jul,
       volume = {897},
       number = {1},
          eid = {93},
        pages = {93},
          doi = {10.3847/1538-4357/ab9a2f},
archivePrefix = {arXiv},
       eprint = {2006.04808},
 primaryClass = {astro-ph.GA},
       adsurl = {https://ui.adsabs.harvard.edu/abs/2020ApJ...897...93B}
}

@article{cassano05,
       author = {{Cassano}, R. and {Brunetti}, G.},
        title = "{Cluster mergers and non-thermal phenomena: a statistical magneto-turbulent model}",
      journal = {\mnras},
         year = 2005,
        month = mar,
       volume = {357},
       number = {4},
        pages = {1313-1329},
          doi = {10.1111/j.1365-2966.2005.08747.x},
archivePrefix = {arXiv},
       eprint = {astro-ph/0412475},
 primaryClass = {astro-ph},
       adsurl = {https://ui.adsabs.harvard.edu/abs/2005MNRAS.357.1313C}
}

@article{cassano10lofar,
       author = {{Cassano}, R. and {Brunetti}, G. and {R{\"o}ttgering}, H.~J.~A. and {Br{\"u}ggen}, M.},
        title = "{Unveiling radio halos in galaxy clusters in the LOFAR era}",
      journal = {\aap},
         year = 2010,
        month = jan,
       volume = {509},
          eid = {A68},
        pages = {A68},
          doi = {10.1051/0004-6361/200913063},
archivePrefix = {arXiv},
       eprint = {0910.2025},
 primaryClass = {astro-ph.CO},
       adsurl = {https://ui.adsabs.harvard.edu/abs/2010A&A...509A..68C}
}

@article{lacey93,
       author = {{Lacey}, Cedric and {Cole}, Shaun},
        title = "{Merger rates in hierarchical models of galaxy formation}",
      journal = {\mnras},
         year = 1993,
        month = jun,
       volume = {262},
       number = {3},
        pages = {627-649},
          doi = {10.1093/mnras/262.3.627},
       adsurl = {https://ui.adsabs.harvard.edu/abs/1993MNRAS.262..627L}
}

@article{fujita03,
       author = {{Fujita}, Yutaka and {Takizawa}, Motokazu and {Sarazin}, Craig L.},
        title = "{Nonthermal Emissions from Particles Accelerated by Turbulence in Clusters of Galaxies}",
      journal = {\apj},
         year = 2003,
        month = feb,
       volume = {584},
       number = {1},
        pages = {190-202},
          doi = {10.1086/345599},
archivePrefix = {arXiv},
       eprint = {astro-ph/0210320},
 primaryClass = {astro-ph},
       adsurl = {https://ui.adsabs.harvard.edu/abs/2003ApJ...584..190F}
}

@article{cassano08revised,
       author = {{Cassano}, R. and {Brunetti}, G. and {Venturi}, T. and {Setti}, G. and {Dallacasa}, D. and {Giacintucci}, S. and {Bardelli}, S.},
        title = "{Revised statistics of radio halos and the reacceleration model}",
      journal = {\aap},
         year = 2008,
        month = mar,
       volume = {480},
       number = {3},
        pages = {687-697},
          doi = {10.1051/0004-6361:20078986},
archivePrefix = {arXiv},
       eprint = {0712.3516},
 primaryClass = {astro-ph},
       adsurl = {https://ui.adsabs.harvard.edu/abs/2008A&A...480..687C}
}

@article{bonafede10,
       author = {{Bonafede}, A. and {Feretti}, L. and {Murgia}, M. and {Govoni}, F. and {Giovannini}, G. and {Dallacasa}, D. and {Dolag}, K. and {Taylor}, G.~B.},
        title = "{The Coma cluster magnetic field from Faraday rotation measures}",
      journal = {\aap},
         year = 2010,
        month = apr,
       volume = {513},
          eid = {A30},
        pages = {A30},
          doi = {10.1051/0004-6361/200913696},
archivePrefix = {arXiv},
       eprint = {1002.0594},
 primaryClass = {astro-ph.CO},
       adsurl = {https://ui.adsabs.harvard.edu/abs/2010A&A...513A..30B}
}

@article{botteon21ant,
       author = {{Botteon}, A. and {Cassano}, R. and {van Weeren}, R.~J. and {Shimwell}, T.~W. and {Bonafede}, A. and {Brggen}, M. and {Brunetti}, G. and {Cuciti}, V. and {Dallacasa}, D. and {de Gasperin}, F. and {Di Gennaro}, G. and {Gastaldello}, F. and {Hoang}, D.~N. and {Rossetti}, M. and {R{\"o}ttgering}, H.~J.~A.},
        title = "{Discovery of a Radio Halo (and Relic) in a M$_{500}$<2{\texttimes}{}10$^{14}$ M$_{☉}$ Cluster}",
      journal = {\apjl},
         year = 2021,
        month = jun,
       volume = {914},
       number = {2},
          eid = {L29},
        pages = {L29},
          doi = {10.3847/2041-8213/ac0636},
archivePrefix = {arXiv},
       eprint = {2105.14025},
 primaryClass = {astro-ph.CO},
       adsurl = {https://ui.adsabs.harvard.edu/abs/2021ApJ...914L..29B}
}

@article{vazza06,
       author = {{Vazza}, F. and {Tormen}, G. and {Cassano}, R. and {Brunetti}, G. and {Dolag}, K.},
        title = "{Turbulent velocity fields in smoothed particle hydrodymanics simulated galaxy clusters: scaling laws for the turbulent energy}",
      journal = {\mnras},
         year = 2006,
        month = jun,
       volume = {369},
       number = {1},
        pages = {L14-L18},
          doi = {10.1111/j.1745-3933.2006.00164.x},
archivePrefix = {arXiv},
       eprint = {astro-ph/0602247},
 primaryClass = {astro-ph},
       adsurl = {https://ui.adsabs.harvard.edu/abs/2006MNRAS.369L..14V}
}

@article{hallman11,
  author = {Hallman, E. J. and Skillman, S. W. and O'Shea, B. W. and Burns, J. O. and Norman, M. L.},
  year = {2011},
  title = {Cosmological simulations of galaxy clusters: turbulent motions and shocks},
  journal = {ApJ},
  volume = {731},
  pages = {L18},
  doi = {10.1088/2041-8205/731/1/L18}
}

@article{cuciti15,
       author = {{Cuciti}, V. and {Cassano}, R. and {Brunetti}, G. and {Dallacasa}, D. and {Kale}, R. and {Ettori}, S. and {Venturi}, T.},
        title = "{Occurrence of radio halos in galaxy clusters. Insight from a mass-selected sample}",
      journal = {\aap},
         year = 2015,
        month = aug,
       volume = {580},
          eid = {A97},
        pages = {A97},
          doi = {10.1051/0004-6361/201526420},
archivePrefix = {arXiv},
       eprint = {1506.03209},
 primaryClass = {astro-ph.CO},
       adsurl = {https://ui.adsabs.harvard.edu/abs/2015A&A...580A..97C}
}

@article{cuciti21b,
       author = {{Cuciti}, V. and {Cassano}, R. and {Brunetti}, G. and {Dallacasa}, D. and {de Gasperin}, F. and {Ettori}, S. and {Giacintucci}, S. and {Kale}, R. and {Pratt}, G.~W. and {van Weeren}, R.~J. and {Venturi}, T.},
        title = "{Radio halos in a mass-selected sample of 75 galaxy clusters. II. Statistical analysis}",
      journal = {\aap},
         year = 2021,
        month = mar,
       volume = {647},
          eid = {A51},
        pages = {A51},
          doi = {10.1051/0004-6361/202039208},
archivePrefix = {arXiv},
       eprint = {2101.01641},
 primaryClass = {astro-ph.CO},
       adsurl = {https://ui.adsabs.harvard.edu/abs/2021A&A...647A..51C}
}

@article{cassano12,
       author = {{Cassano}, R. and {Brunetti}, G. and {Norris}, R.~P. and {R{\"o}ttgering}, H.~J.~A. and {Johnston-Hollitt}, M. and {Trasatti}, M.},
        title = "{Radio halos in future surveys in the radio continuum}",
      journal = {\aap},
         year = 2012,
        month = dec,
       volume = {548},
          eid = {A100},
        pages = {A100},
          doi = {10.1051/0004-6361/201220018},
archivePrefix = {arXiv},
       eprint = {1210.1020},
 primaryClass = {astro-ph.CO},
       adsurl = {https://ui.adsabs.harvard.edu/abs/2012A&A...548A.100C}
}

@article{bonafede2014,
       author = {{Bonafede}, A. and {Intema}, H.~T. and {Bruggen}, M. and {Russell}, H.~R. and {Ogrean}, G. and {Basu}, K. and {Sommer}, M. and {van Weeren}, R.~J. and {Cassano}, R. and {Fabian}, A.~C. and {Rottgering}, H.~J.~A.},
        title = "{A giant radio halo in the cool core cluster CL1821+643.}",
      journal = {\mnras},
         year = 2014,
        month = oct,
       volume = {444},
        pages = {L44-L48},
          doi = {10.1093/mnrasl/slu110},
archivePrefix = {arXiv},
       eprint = {1407.4801},
 primaryClass = {astro-ph.CO},
       adsurl = {https://ui.adsabs.harvard.edu/abs/2014MNRAS.444L..44B}
}

@article{savini2018,
       author = {{Savini}, F. and {Bonafede}, A. and {Br{\"u}ggen}, M. and {van Weeren}, R. and {Brunetti}, G. and {Intema}, H. and {Botteon}, A. and {Shimwell}, T. and {Wilber}, A. and {Rafferty}, D. and {Giacintucci}, S. and {Cassano}, R. and {Cuciti}, V. and {de Gasperin}, F. and {R{\"o}ttgering}, H. and {Hoeft}, M. and {White}, G.},
        title = "{First evidence of diffuse ultra-steep-spectrum radio emission surrounding the cool core of a cluster}",
      journal = {\mnras},
         year = 2018,
        month = aug,
       volume = {478},
       number = {2},
        pages = {2234-2242},
          doi = {10.1093/mnras/sty1125},
archivePrefix = {arXiv},
       eprint = {1805.01900},
 primaryClass = {astro-ph.GA},
       adsurl = {https://ui.adsabs.harvard.edu/abs/2018MNRAS.478.2234S}
}

@article{brunetti2001,
       author = {{Brunetti}, G. and {Setti}, G. and {Feretti}, L. and {Giovannini}, G.},
        title = "{Particle injection and reacceleration in clusters of galaxies and the EUV excess: the case of Coma}",
      journal = {\na},
         year = 2001,
        month = feb,
       volume = {6},
       number = {1},
        pages = {1-15},
          doi = {10.1016/S1384-1076(00)00048-8},
archivePrefix = {arXiv},
       eprint = {astro-ph/0011301},
 primaryClass = {astro-ph},
       adsurl = {https://ui.adsabs.harvard.edu/abs/2001NewA....6....1B}
}

@article{petrosian2001,
   author = {{Petrosian}, V.},
    title = "{On the Nonthermal Emission and Acceleration of Electrons in Coma and Other Clusters of Galaxies}",
  journal = {\apj},
   eprint = {astro-ph/0101145},
     year = 2001,
    month = aug,
   volume = 557,
    pages = {560-572},
      doi = {10.1086/321557},
   adsurl = {http://adsabs.harvard.edu/abs/2001ApJ...557..560P}
}

@article{brunetti2016,
       author = {{Brunetti}, G. and {Lazarian}, A.},
        title = "{Stochastic reacceleration of relativistic electrons by turbulent reconnection: a mechanism for cluster-scale radio emission?}",
      journal = {\mnras},
         year = 2016,
        month = may,
       volume = {458},
       number = {3},
        pages = {2584-2595},
          doi = {10.1093/mnras/stw496},
archivePrefix = {arXiv},
       eprint = {1603.00458},
 primaryClass = {astro-ph.HE},
       adsurl = {https://ui.adsabs.harvard.edu/abs/2016MNRAS.458.2584B}
}

@article{cassano2006,
       author = {{Cassano}, R. and {Brunetti}, G. and {Setti}, G.},
        title = "{Statistics of giant radio haloes from electron reacceleration models}",
      journal = {\mnras},
         year = 2006,
        month = jul,
       volume = {369},
       number = {4},
        pages = {1577-1595},
          doi = {10.1111/j.1365-2966.2006.10423.x},
archivePrefix = {arXiv},
       eprint = {astro-ph/0604103},
 primaryClass = {astro-ph},
       adsurl = {https://ui.adsabs.harvard.edu/abs/2006MNRAS.369.1577C}
}

@INPROCEEDINGS{cassano2015,
       author = {{Cassano}, R. and {Bernardi}, G. and {Brunetti}, G. and {Br{\"u}ggen}, M. and {Clarke}, T. and {Dallacasa}, D. and {Dolag}, K. and {Ettori}, S. and {Giacintucci}, S. and {Giocoli}, C. and {Gitti}, M. and {Johnston-Hollitt}, M. and {Kale}, R. and {Markevich}, M. and {Norris}, R. and {Pommier}, M.~P. and {Pratt}, G. and {Rottgering}, H.~J.~A. and {Venturi}, T.},
        title = "{Cluster Radio Halos at the crossroads between astrophysics and cosmology in the SKA era}",
    booktitle = {Advancing Astrophysics with the Square Kilometre Array (AASKA14)},
         year = 2015,
        month = apr,
          eid = {73},
        pages = {73},
          doi = {10.22323/1.215.0073},
archivePrefix = {arXiv},
       eprint = {1412.5940},
 primaryClass = {astro-ph.CO},
       adsurl = {https://ui.adsabs.harvard.edu/abs/2015aska.confE..73C}
}

@article{donnert2013,
       author = {{Donnert}, J. and {Dolag}, K. and {Brunetti}, G. and {Cassano}, R.},
        title = "{Rise and fall of radio haloes in simulated merging galaxy clusters}",
      journal = {\mnras},
         year = 2013,
        month = Mar,
       volume = {429},
        pages = {3564-3569},
          doi = {10.1093/mnras/sts628},
archivePrefix = {arXiv},
       eprint = {1211.3337},
 primaryClass = {astro-ph.CO},
       adsurl = {https://ui.adsabs.harvard.edu/#abs/2013MNRAS.429.3564D}
}

@article{dennison1980,
       author = {{Dennison}, B.},
        title = "{Formation of radio halos in clusters of galaxies from cosmic-ray protons.}",
      journal = {\apjl},
         year = 1980,
        month = aug,
       volume = {239},
        pages = {L93-L96},
          doi = {10.1086/183300},
       adsurl = {https://ui.adsabs.harvard.edu/abs/1980ApJ...239L..93D}
}

@article{blasi1999,
       author = {{Blasi}, Pasquale and {Colafrancesco}, Sergio},
        title = "{Cosmic rays, radio halos and nonthermal X-ray emission in clusters of galaxies}",
      journal = {Astroparticle Physics},
         year = 1999,
        month = nov,
       volume = {12},
       number = {3},
        pages = {169-183},
          doi = {10.1016/S0927-6505(99)00079-1},
archivePrefix = {arXiv},
       eprint = {astro-ph/9905122},
 primaryClass = {astro-ph},
       adsurl = {https://ui.adsabs.harvard.edu/abs/1999APh....12..169B}
}

@article{ackermann2016,
       author = {{Ackermann}, M. and {Albert}, A. and {Atwood}, W.~B. and {Baldini}, L. and {Ballet}, J. and {Barbiellini}, G. and {Bastieri}, D. and {Bellazzini}, R. and {Bissaldi}, E. and {Bloom}, E.~D. and {Bonino}, R. and {Brandt}, T.~J. and {Bregeon}, J. and {Bruel}, P. and {Buehler}, R. and {Caliandro}, G.~A. and {Cameron}, R.~A. and {Caragiulo}, M. and {Caraveo}, P.~A. and {Cavazzuti}, E. and {Cecchi}, C. and {Charles}, E. and {Chekhtman}, A. and {Chiang}, J. and {Chiaro}, G. and {Ciprini}, S. and {Cohen-Tanugi}, J. and {Cutini}, S. and {D'Ammando}, F. and {de Angelis}, A. and {de Palma}, F. and {Desiante}, R. and {Digel}, S.~W. and {Drell}, P.~S. and {Favuzzi}, C. and {Ferrara}, E.~C. and {Focke}, W.~B. and {Franckowiak}, A. and {Fusco}, P. and {Gargano}, F. and {Gasparrini}, D. and {Giglietto}, N. and {Giordano}, F. and {Godfrey}, G. and {Grenier}, I.~A. and {Grondin}, M.-H. and {Guillemot}, L. and {Guiriec}, S. and {Harding}, A.~K. and {Hill}, A.~B. and {Horan}, D. and {J{\'o}hannesson}, G. and {Kn{\"o}dlseder}, J. and {Kuss}, M. and {Larsson}, S. and {Latronico}, L. and {Li}, J. and {Li}, L. and {Longo}, F. and {Loparco}, F. and {Lubrano}, P. and {Maldera}, S. and {Martin}, P. and {Mayer}, M. and {Mazziotta}, M.~N. and {Michelson}, P.~F. and {Mizuno}, T. and {Monzani}, M.~E. and {Morselli}, A. and {Murgia}, S. and {Nuss}, E. and {Ohsugi}, T. and {Orienti}, M. and {Orlando}, E. and {Ormes}, J.~F. and {Paneque}, D. and {Pesce-Rollins}, M. and {Piron}, F. and {Pivato}, G. and {Porter}, T.~A. and {Rain{\`o}}, S. and {Rando}, R. and {Razzano}, M. and {Reimer}, A. and {Reimer}, O. and {Romani}, R.~W. and {S{\'a}nchez-Conde}, M. and {Schulz}, A. and {Sgr{\`o}}, C. and {Siskind}, E.~J. and {Smith}, D.~A. and {Spada}, F. and {Spandre}, G. and {Spinelli}, P. and {Suson}, D.~J. and {Takahashi}, H. and {Thayer}, J.~B. and {Tibaldo}, L. and {Torres}, D.~F. and {Tosti}, G. and {Troja}, E. and {Vianello}, G. and {Wood}, M. and {Zimmer}, S.},
        title = "{Deep view of the Large Magellanic Cloud with six years of Fermi-LAT observations}",
      journal = {\aap},
         year = 2016,
        month = feb,
       volume = {586},
          eid = {A71},
        pages = {A71},
          doi = {10.1051/0004-6361/201526920},
archivePrefix = {arXiv},
       eprint = {1509.06903},
 primaryClass = {astro-ph.HE},
       adsurl = {https://ui.adsabs.harvard.edu/abs/2016A&A...586A..71A}
}

@article{brunetti2017,
       author = {{Brunetti}, G. and {Zimmer}, S. and {Zandanel}, F.},
        title = "{Relativistic protons in the Coma galaxy cluster: first gamma-ray constraints ever on turbulent reacceleration}",
      journal = {\mnras},
         year = 2017,
        month = dec,
       volume = {472},
       number = {2},
        pages = {1506-1525},
          doi = {10.1093/mnras/stx2092},
archivePrefix = {arXiv},
       eprint = {1707.02085},
 primaryClass = {astro-ph.HE},
       adsurl = {https://ui.adsabs.harvard.edu/abs/2017MNRAS.472.1506B}
}

@article{adam2021,
       author = {{Adam}, R. and {Goksu}, H. and {Brown}, S. and {Rudnick}, L. and {Ferrari}, C.},
        title = "{{\ensuremath{\gamma}}-ray detection toward the Coma cluster with Fermi-LAT: Implications for the cosmic ray content in the hadronic scenario}",
      journal = {\aap},
         year = 2021,
        month = apr,
       volume = {648},
          eid = {A60},
        pages = {A60},
          doi = {10.1051/0004-6361/202039660},
archivePrefix = {arXiv},
       eprint = {2102.02251},
 primaryClass = {astro-ph.HE},
       adsurl = {https://ui.adsabs.harvard.edu/abs/2021A&A...648A..60A}
}

@techreport{braun2014,
  author       = {Braun, R.},
  title        = {SKA1 Imaging Science Performance},
  institution  = {SKA Organisation},
  number       = {SKA-TEL-SKO-DD-XXX, Rev. A Draft 2},
  year         = {2014},
  url          = {https://indico.skatelescope.org/event/270/attachments/1919/2406/SKA1_Science_PerformanceRevA_draft2.pdf},
  note         = {Technical report}
}

@article{braun2019,
  author       = {Braun, R. and {et al.}},
  title        = {Anticipated performance of SKA1‑LOW and SKA1‑MID surveys},
  journal      = {Proceedings of the International Astronomical Union},
  year         = {2019},
  volume       = {15},
  pages        = {362–365},
  note         = {SKA1 survey sensitivity and capabilities}  
}

@article{cassano06,
author = {Cassano, Rossella and Brunetti, Gianfranco and Setti, Giancarlo},
doi = {10.1111/j.1365-2966.2006.10423.x},
journal = {MNRAS},
pages = {1577--1595},
title = {{Statistics of giant radio haloes from electron reacceleration models}},
volume = {369},
year = {2006}
}

@ARTICLE{XRISM_2319_2025,
       author = {{XRISM Collaboration} and {Audard}, Marc and {Awaki}, Hisamitsu and {Ballhausen}, Ralf and {Bamba}, Aya and {Behar}, Ehud and {Boissay-malaquin}, Rozenn and {Brenneman}, Laura and {Brown}, Gregory V. and {Corrales}, Lia and {Costantini}, Elisa and {Cumbee}, Renata and {Diaz Trigo}, Maria and {Done}, Chris and {Dotani}, Tadayasu and {Ebisawa}, Ken and {Eckart}, Megan E. and {Eckert}, Dominique and {Eguchi}, Satoshi and {Enoto}, Teruaki and {Ezoe}, Yuichiro and {Foster}, Adam and {Fujimoto}, Ryuichi and {Fujita}, Yutaka and {Fukazawa}, Yasushi and {Fukushima}, Kotaro and {Furuzawa}, Akihiro and {Gallo}, Luigi and {Garc{\'\i}a}, Javier and {Gu}, Liyi and {Guainazzi}, Matteo and {Hagino}, Kouichi and {Hamaguchi}, Kenji and {Hatsukade}, Isamu and {Hayashi}, Katsuhiro and {Hayashi}, Takayuki and {Hell}, Natalie and {Hodges-kluck}, Edmund and {Hornschemeier}, Ann and {Ichinohe}, Yuto and {Ishi}, Daiki and {Ishida}, Manabu and {Ishikawa}, Kumi and {Ishisaki}, Yoshitaka and {Kaastra}, Jelle and {Kallman}, Timothy and {Kara}, Erin and {Katsuda}, Satoru and {Kanemaru}, Yoshiaki and {Kelley}, Richard and {Kilbourne}, Caroline and {Kitamoto}, Shunji and {Kobayashi}, Shogo B. and {Kohmura}, Takayoshi and {Kubota}, Aya and {Leutenegger}, Maurice and {Loewenstein}, Michael and {Maeda}, Yoshitomo and {Markevitch}, Maxim and {Matsumoto}, Hironori and {Matsushita}, Kyoko and {Mccammon}, Dan and {Mcnamara}, Brian and {Mernier}, Francois and {Miller}, Eric and {Miller}, Jon and {Mitsuishi}, Ikuyuki and {Mizumoto}, Misaki and {Mizuno}, Tsunefumi and {Mori}, Koji and {Mukai}, Koji and {Murakami}, Hiroshi and {Mushotzky}, Richard and {Nakajima}, Hiroshi and {Nakazawa}, Kazuhiro and {Ness}, Jan-uwe and {Nobukawa}, Kumiko and {Nobukawa}, Masayoshi and {Noda}, Hirofumi and {Odaka}, Hirokazu and {Ogawa}, Shoji and {Ogorzalek}, Anna and {Okajima}, Takashi and {Ota}, Naomi and {Paltani}, Stephane and {Petre}, Robert and {Plucinsky}, Paul and {Porter}, Frederick and {Pottschmidt}, Katja and {Sato}, Kosuke and {Sato}, Toshiki and {Sawada}, Makoto and {Seta}, Hiromi and {Shidatsu}, Megumi and {Simionescu}, Aurora and {Smith}, Randall and {Suzuki}, Hiromasa and {Szymkowiak}, Andrew and {Takahashi}, Hiromitsu and {Takeo}, Mai and {Tamagawa}, Toru and {Tamura}, Keisuke and {Tanaka}, Takaaki and {Tanimoto}, Atsushi and {Tashiro}, Makoto and {Terada}, Yukikatsu and {Terashima}, Yuichi and {Tsuboi}, Yohko and {Tsujimoto}, Masahiro and {Tsunemi}, Hiroshi and {Tsuru}, Takeshi and {Uchida}, Hiroyuki and {Uchida}, Nagomi and {Uchida}, Yuusuke and {Uchiyama}, Hideki and {Ueda}, Yoshihiro and {Uno}, Shinichiro and {Vink}, Jacco and {Watanabe}, Shin and {Williams}, Brian J. and {Yamada}, Satoshi and {Yamada}, Shinya and {Yamaguchi}, Hiroya and {Yamaoka}, Kazutaka and {Yamasaki}, Noriko and {Yamauchi}, Makoto and {Yamauchi}, Shigeo and {Yaqoob}, Tahir and {Yoneyama}, Tomokage and {Yoshida}, Tessei and {Yukita}, Mihoko and {Zhuravleva}, Irina and {Seppi}, Riccardo and {Aihara}, Itsuki and {Omiya}, Yuki},
        title = "{XRISM/Resolve View of Abell 2319: Turbulence, Sloshing, and ICM Dynamics}",
      journal = {arXiv e-prints},
         year = 2025,
        month = aug,
          eid = {arXiv:2508.05067},
        pages = {arXiv:2508.05067},
          doi = {10.48550/arXiv.2508.05067},
archivePrefix = {arXiv},
       eprint = {2508.05067},
 primaryClass = {astro-ph.HE},
       adsurl = {https://ui.adsabs.harvard.edu/abs/2025arXiv250805067X}
}

@ARTICLE{XRISM_A2029_2025,
       author = {{XRISM Collaboration} and {Audard}, Marc and {Awaki}, Hisamitsu and {Ballhausen}, Ralf and {Bamba}, Aya and {Behar}, Ehud and {Boissay-Malaquin}, Rozenn and {Brenneman}, Laura and {Brown}, Gregory V. and {Corrales}, Lia and {Costantini}, Elisa and {Cumbee}, Renata and {Diaz Trigo}, Maria and {Done}, Chris and {Dotani}, Tadayasu and {Ebisawa}, Ken and {Eckart}, Megan E. and {Eckert}, Dominique and {Eguchi}, Satoshi and {Enoto}, Teruaki and {Ezoe}, Yuichiro and {Foster}, Adam and {Fujimoto}, Ryuichi and {Fujita}, Yutaka and {Fukazawa}, Yasushi and {Fukushima}, Kotaro and {Furuzawa}, Akihiro and {Gallo}, Luigi and {Garc{\'\i}a}, Javier A. and {Gu}, Liyi and {Guainazzi}, Matteo and {Hagino}, Kouichi and {Hamaguchi}, Kenji and {Hatsukade}, Isamu and {Hayashi}, Katsuhiro and {Hayashi}, Takayuki and {Hell}, Natalie and {Hodges-Kluck}, Edmund and {Hornschemeier}, Ann and {Ichinohe}, Yuto and {Ishida}, Manabu and {Ishikawa}, Kumi and {Ishisaki}, Yoshitaka and {Kaastra}, Jelle and {Kallman}, Timothy and {Kara}, Erin and {Katsuda}, Satoru and {Kanemaru}, Yoshiaki and {Kelley}, Richard and {Kilbourne}, Caroline and {Kitamoto}, Shunji and {Kobayashi}, Shogo and {Kohmura}, Takayoshi and {Kubota}, Aya and {Leutenegger}, Maurice and {Loewenstein}, Michael and {Maeda}, Yoshitomo and {Markevitch}, Maxim and {Matsumoto}, Hironori and {Matsushita}, Kyoko and {McCammon}, Dan and {McNamara}, Brian and {Mernier}, Fran{\c{c}}ois and {Miller}, Eric D. and {Miller}, Jon M. and {Mitsuishi}, Ikuyuki and {Mizumoto}, Misaki and {Mizuno}, Tsunefumi and {Mori}, Koji and {Mukai}, Koji and {Murakami}, Hiroshi and {Mushotzky}, Richard and {Nakajima}, Hiroshi and {Nakazawa}, Kazuhiro and {Ness}, Jan-Uwe and {Nobukawa}, Kumiko and {Nobukawa}, Masayoshi and {Noda}, Hirofumi and {Odaka}, Hirokazu and {Ogawa}, Shoji and {Ogorzalek}, Anna and {Okajima}, Takashi and {Ota}, Naomi and {Paltani}, Stephane and {Petre}, Robert and {Plucinsky}, Paul and {Porter}, Frederick S. and {Pottschmidt}, Katja and {Sato}, Kosuke and {Sato}, Toshiki and {Sawada}, Makoto and {Seta}, Hiromi and {Shidatsu}, Megumi and {Simionescu}, Aurora and {Smith}, Randall and {Suzuki}, Hiromasa and {Szymkowiak}, Andrew and {Takahashi}, Hiromitsu and {Takeo}, Mai and {Tamagawa}, Toru and {Tamura}, Keisuke and {Tanaka}, Takaaki and {Tanimoto}, Atsushi and {Tashiro}, Makoto and {Terada}, Yukikatsu and {Terashima}, Yuichi and {Tsuboi}, Yohko and {Tsujimoto}, Masahiro and {Tsunemi}, Hiroshi and {Tsuru}, Takeshi and {Uchida}, Hiroyuki and {Uchida}, Nagomi and {Uchida}, Yuusuke and {Uchiyama}, Hideki and {Ueda}, Yoshihiro and {Uno}, Shinichiro and {Vink}, Jacco and {Watanabe}, Shin and {Williams}, Brian J. and {Yamada}, Satoshi and {Yamada}, Shinya and {Yamaguchi}, Hiroya and {Yamaoka}, Kazutaka and {Yamasaki}, Noriko and {Yamauchi}, Makoto and {Yamauchi}, Shigeo and {Yaqoob}, Tahir and {Yoneyama}, Tomokage and {Yoshida}, Tessei and {Yukita}, Mihoko and {Zhuravleva}, Irina and {Bartalesi}, Tommaso and {Ettori}, Stefano and {Kosarzycki}, Roman and {Lovisari}, Lorenzo and {Rose}, Tom and {Sarkar}, Arnab and {Sun}, Ming and {Tamhane}, Prathamesh},
        title = "{XRISM Reveals Low Nonthermal Pressure in the Core of the Hot, Relaxed Galaxy Cluster A2029}",
      journal = {\apjl},
         year = 2025,
        month = mar,
       volume = {982},
       number = {1},
          eid = {L5},
        pages = {L5},
          doi = {10.3847/2041-8213/ada7cd},
archivePrefix = {arXiv},
       eprint = {2501.05514},
 primaryClass = {astro-ph.HE},
       adsurl = {https://ui.adsabs.harvard.edu/abs/2025ApJ...982L...5X}
}

@ARTICLE{Vazza_2025,
       author = {{Vazza}, F. and {Brunetti}, G.},
        title = "{On the interpretation of XRISM X-ray measurements of turbulence in the intracluster medium: a comparison with cosmological simulations}",
      journal = {arXiv e-prints},
         year = 2025,
        month = jul,
          eid = {arXiv:2507.04727},
        pages = {arXiv:2507.04727},
          doi = {10.48550/arXiv.2507.04727},
archivePrefix = {arXiv},
       eprint = {2507.04727},
 primaryClass = {astro-ph.CO},
       adsurl = {https://ui.adsabs.harvard.edu/abs/2025arXiv250704727V}
 }

@article{rajpurohit21A2744,
  author = {Rajpurohit, K. et al.},
  year = {2021},
  title = {Revealing extreme steep-spectrum radio halo emission in Abell 2744 with MeerKAT},
  journal = {Astronomy \& Astrophysics},
  volume = {655},
  pages = {A30},
  doi = {10.1051/0004-6361/202141499}
}

@ARTICLE{Balboni24,
       author = {{Balboni}, M. and {Gastaldello}, F. and {Bonafede}, A. and {Botteon}, A. and {Bartalucci}, I. and {Bourdin}, H. and {Brunetti}, G. and {Cassano}, R. and {De Grandi}, S. and {De Luca}, F. and {Ettori}, S. and {Ghizzardi}, S. and {Gitti}, M. and {Iqbal}, A. and {Johnston-Hollitt}, M. and {Lovisari}, L. and {Mazzotta}, P. and {Molendi}, S. and {Pointecouteau}, E. and {Pratt}, G.~W. and {Riva}, G. and {Rossetti}, M. and {Rottgering}, H. and {Sereno}, M. and {van Weeren}, R.~J. and {Venturi}, T. and {Veronesi}, I.},
        title = "{CHEX-MATE: A LOFAR pilot X-ray - radio study on five radio halo clusters}",
      journal = {\aap},
         year = 2024,
        month = jun,
       volume = {686},
          eid = {A5},
        pages = {A5},
          doi = {10.1051/0004-6361/202347965},
archivePrefix = {arXiv},
       eprint = {2402.18654},
 primaryClass = {astro-ph.CO},
       adsurl = {https://ui.adsabs.harvard.edu/abs/2024A&A...686A...5B}
}

@article{feretti2012,
       author = {{Feretti}, Luigina and {Giovannini}, Gabriele and {Govoni}, Federica and {Murgia}, Matteo},
        title = "{Clusters of galaxies: observational properties of the diffuse radio emission}",
      journal = {\aapr},
         year = 2012,
        month = may,
       volume = {20},
          eid = {54},
        pages = {54},
          doi = {10.1007/s00159-012-0054-z},
archivePrefix = {arXiv},
       eprint = {1205.1919},
 primaryClass = {astro-ph.CO},
       adsurl = {https://ui.adsabs.harvard.edu/abs/2012A&ARv..20...54F}
}

\end{document}